\documentclass[a4paper,fleqn,usenatbib]{mnras}

\usepackage[T1]{fontenc}
\usepackage{ae,aecompl}
\usepackage{etoolbox}
\makeatletter
\patchcmd\@combinedblfloats{\box\@outputbox}{\unvbox\@outputbox}{}{%
   \errmessage{\noexpand\@combinedblfloats could not be patched}%
}%
 \makeatother
\usepackage{graphicx}	% Including figure files
\usepackage{amsmath}	% Advanced maths commands
\usepackage{amssymb}	% Extra maths symbols
\usepackage{comment}
\newcommand {\kms} {\,{\rm km\,s}^{-1}}

\newcommand {\mo}{{\rm M}_\odot}

\newcommand {\moyr}{\,{\rm M_\odot\,\rm yr}^{-1}}

\def\hi{\ifmmode{\rm HI}\else{H\/{\sc i}}\fi} 
\def\oi{\ifmmode{\rm OI}\else{O\/{\sc i}}\fi} 
\def\ci{\ifmmode{\rm CI}\else{C\/{\sc i}}\fi} 
 \def\cii{\ifmmode{\rm CII}\else{C\/{\sc ii}}\fi}

\newcommand{\aref}[1]{\hyperref[#1]{Appendix~\ref{#1}}}

\defcitealias{Armillotta+19}{Paper I}

\title[Life cycle of the CMZ]{The Life Cycle of the Central Molecular Zone. II: Distribution of atomic and molecular gas tracers}

\author[L. Armillotta et al.]{Lucia Armillotta$^{1,2}$\thanks{E-mail: lucia.armillotta@princeton.edu}, Mark R. Krumholz$^{1,3}$, and  Enrico M. Di Teodoro$^{1,4}$\\
$^{1}$Research School of Astronomy and Astrophysics - The Australian National University, Canberra, ACT, 2611, Australia\\
$^{2}$Department of Astrophysical Sciences, Princeton University, Princeton, NJ 08544, USA\\
$^{3}$Centre of Excellence for Astronomy in Three Dimensions (ASTRO-3D), Australia\\
$^{4}$Department of Physics \& Astronomy, Johns Hopkins University, Baltimore, MD 21218, USA }
\date{Accepted XXX. Received YYY; in original form ZZZ}

\pubyear{2019}

\begin{document}
\label{firstpage}
\pagerange{\pageref{firstpage}--\pageref{lastpage}}
\maketitle

\begin{abstract}
We use the hydrodynamical simulation of our inner Galaxy presented in \citet{Armillotta+19} to study the gas distribution and kinematics within the CMZ. We use a resolution high enough to capture the gas emitting in dense molecular tracers such as NH$_3$ and HCN, and simulate a time window of 50 Myr, long enough to capture phases during which the CMZ experiences both quiescent and intense star formation. We then post-process the simulated CMZ to calculate its spatially-dependent chemical and thermal state, producing synthetic emission datacubes and maps of both \hi\ and the molecular gas tracers CO, NH$_3$ and HCN. We show that, as viewed from Earth, gas in the CMZ is distributed mainly in two parallel and elongated features extending from positive longitudes and velocities to negative longitudes and velocities. The molecular gas emission within these two streams is not uniform, and it is mostly associated to the region where gas flowing towards the Galactic Center through the dust lanes collides with gas orbiting within the ring. Our simulated data cubes reproduce a number of features found in the observed CMZ. However, some discrepancies emerge when we use our results to interpret the position of individual molecular clouds. Finally, we show that, when the CMZ is near a period of intense star formation, the ring is mostly fragmented as a consequence of supernova feedback, and the bulk of the emission comes from star-forming molecular clouds. This correlation between morphology and star formation rate should be detectable in observations of extragalactic CMZs. 
\end{abstract}


%%%%%%%%%%%%%%%%%%%%%%%%%%%%%%%%%%%%%%%%%%%%%%%%%%

%%%%%%%%%%%%%%%%% BODY OF PAPER %%%%%%%%%%%%%%%%%%

\section{Introduction}
\label{Introduction}

The innermost few hundred parsecs region of the Milky Way, the so-called ``Central Molecular Zone'' \citep[CMZ,][]{Morris&Serabyn96}, contains a large reservoir of molecular gas \citep[$M_\mathrm{CMZ} \sim 2-7\times10^7\, \mo$, e.g.][]{Ferriere+07, Molinari+11, Longmore+13a} with properties substantially different from those observed in the Galactic disc. Most of this gas exhibits column and volume densities $\sim 2$ orders of magnitude larger than those measured elsewhere in the disc \citep[$n_\mathrm{CMZ}\sim 10^4$~cm$^{-3}$, e.g.][]{Kruijssen&Longmore13, Kruijssen+14, Battersby+17}, warmer temperatures \citep[$T_\mathrm{CMZ}\sim 25-200$~K, e.g.][]{Ao+13, Ginsburg+16, Krieger+17} and higher level of turbulence \citep[e.g.][]{Shetty+12, Federrath+16, Henshaw+16}. Despite the high gas densities, the star formation rate (SFR) in the CMZ is at least one order of magnitude lower than what would be expected if dense gas in the CMZ were to form stars at the same rate as comparably dense gas further out in the disc \citep[$\mathrm{SFR}_\mathrm{CMZ} \sim 0.04 - 0.15 \moyr$, e.g.][]{Yusef-Zadeh+09, Longmore+13a, Barnes+17}.

In the last decades, the gas distribution in the CMZ has been extensively studied with the ultimate goal of understanding the characteristics of this extreme environment. Observations of molecular gas tracers, including CO \citep[e.g.][]{Dame+01, Oka+07}, NH$_3$ \citep[e.g.][]{Purcell+12, Krieger+17}, HCN \citep[][]{Jones+12}, and dust tracers \citep[e.g.][]{Rodriguez+04, Molinari+11} have shown the presence of kinematically coherent structures mostly located within an angular distance of $1^{\circ}$ from the Galactic center (corresponding to a physical distance of $\sim150$~pc). Despite the increasing availability of high-resolution data, a detailed characterization of the 3D geometry of the CMZ is still lacking. In fact, our view of the Milky Way through the disk makes extremely difficult to infer the distance of individual gas components from the Sun. Several different interpretations of the gas distribution in the CMZ have been proposed so far, including two spiral arms \citep[e.g.][]{Sofue95, Sawada+04, Ridley+17}, a closed elliptical orbit \citep{Molinari+11} and an open stream \citep{Kruijssen+15}. Recently, \citet{Henshaw+16} performed a detailed analysis of the molecular gas kinematics in the CMZ, concluding that, among the three mentioned models, the closed elliptical orbit is not able to reproduce the observed data. On the other hand, to distinguish between the other two models, they conclude that proper motion observations are required. The detection of prominent molecular clouds, such as the $20\, \kms$ and $50\,\kms$ clouds and the dust ridge clouds, in absorption at mid-infrared wavelengths \citep[][]{Molinari+11} appears to favour the open stream model, which places all these clouds in front of the Galactic Center. Further constraints on the real position of these clouds with respect to the Galactic Center are expected to come from measurements of their X-ray emission as a function of time, assuming that this emission is the reflection of X-rays generated by the central massive black hole Sgr A* during a past outburst \citep[e.g.][]{Clavel+13,Churazov+17}.

In the absence of direct observational determinations, realistic simulations of Milky Way-like galaxies can help in interpreting the gas distribution and kinematics within the CMZ. In the first paper of this series \citep[][hereafter \citetalias{Armillotta+19}]{Armillotta+19}, we presented a study of the gas cycle and star formation history within the CMZ through hydrodynamical simulations of the inner region (4.5 kpc from the Galactic centre) of our Galaxy. In agreement with similar computational works \citep[e.g.][]{Shin+17, Sormani+17, Seo+19}, we found that most of the gas in the CMZ is located in a dense ring-like structure. Such a ring forms in response  to the presence of a bar in our Galaxy. In a non-axysimmetric gravitational potential there are two possible families of closed stable orbits: $x_1$ orbits, i.e. orbits elongated parallel to the bar major axis, and $x_2$ orbits, i.e. orbits elongated parallel to the bar minor axis \citep[][see also \citealt{Kim&Stone12, Sormani+15, Li+16}]{Binney+91}. Gas in the outer parts of the bar slowly drifts towards the Galactic centre following a sequence of $x_1$ orbits. These orbits are more and more elongated as the Galactic centre is approached. When they become self-intersecting, gas is shocked inwards to $x_2$ orbits and nearly settles into a high-density ring. Depending on local conditions of star formation and stellar feedback, gas may locally depart from $x_2$ orbits, making the overall gas distribution asymmetric. In \citetalias{Armillotta+19}, we showed that star formation mostly takes place within the dense ring-like structure, and that is characterised by oscillatory cycles of burst/quenching. Our simulations suggest that the current CMZ is near a minimum of a star formation cycle, in agreement with the findings of earlier analytic and one-dimensional dynamical models by \citet{Kruijssen+14}, \citet{Krumholz+15a}, and \citet{Krumholz+17}.

In this second work, we re-simulate part of the simulation described in \citetalias{Armillotta+19} at higher resolution. We then perform detailed observational and chemical post-processing of this new simulation and compare the gas distribution within the CMZ with real observational data. Our goal is three-fold. First, we seek to determine whether our simulation, which reproduces many bulk properties of the observed CMZ quite well, is also able to match the detailed structures found in line maps; in particular, we wish to verify if our hypothesis that the CMZ is currently near a minimum of its star formation cycle is consistent with the available data on gas distribution and kinematics. Second, we seek to use the simulations to understand how structures in the position-position-velocity space to which we have observational access map onto structures in physical space. Third, we seek to understand how the observable characteristics of CMZs change over the burst/quench cycle, with the goal of predicting what might be detectable in observations of external galaxies that are at different parts of their cycle than the Milky Way. The paper is organised as follows. In \autoref{Data}, we provide a short review of the main observational features of the CMZ to which we will compare our simulation. In \autoref{Methods}, we describe the main features of the simulation and the details of our chemical and observational post-processing technique. In \autoref{Results}, we analyse the synthetic data produced from the simulation. In \autoref{Comparison}, we compare synthetic and observational data and discuss our results in relation to previous works. Finally, in \autoref{Conclusions}, we summarise our main findings.

\section{Observational data}
\label{Data}

\begin{figure}
\includegraphics[width=0.49\textwidth]{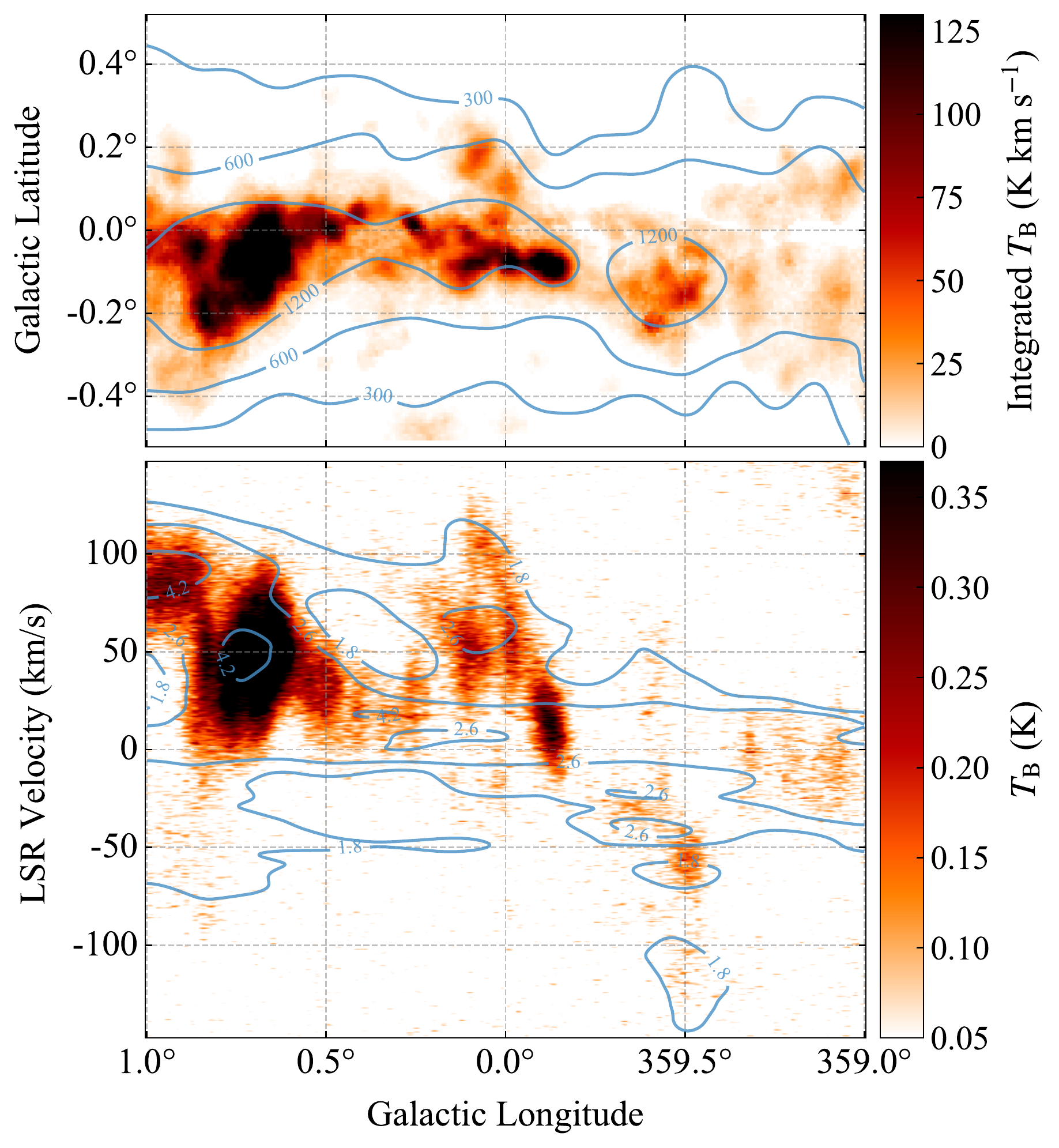}
\caption{ Distribution of NH$_\mathrm{3}$~(1,1) \citep[heat color-scale,][]{Walsh+11,Purcell+12} and CO~($J=1-0$) \citep[blue contours,][]{Bitran+97,Dame+01} emission in the central $2^{\circ}$ of the Galaxy. The \textit{upper panel} shows the integrated brightness temperature map over a velocity range $\pm 350 \kms$, while the \textit{lower panel} shows the longitude-velocity diagrams averaged over latitude $\pm 0.\!\!^{\circ}5$.}
\label{NH3-CO}
\end{figure}

\begin{figure}
\includegraphics[width=0.47\textwidth]{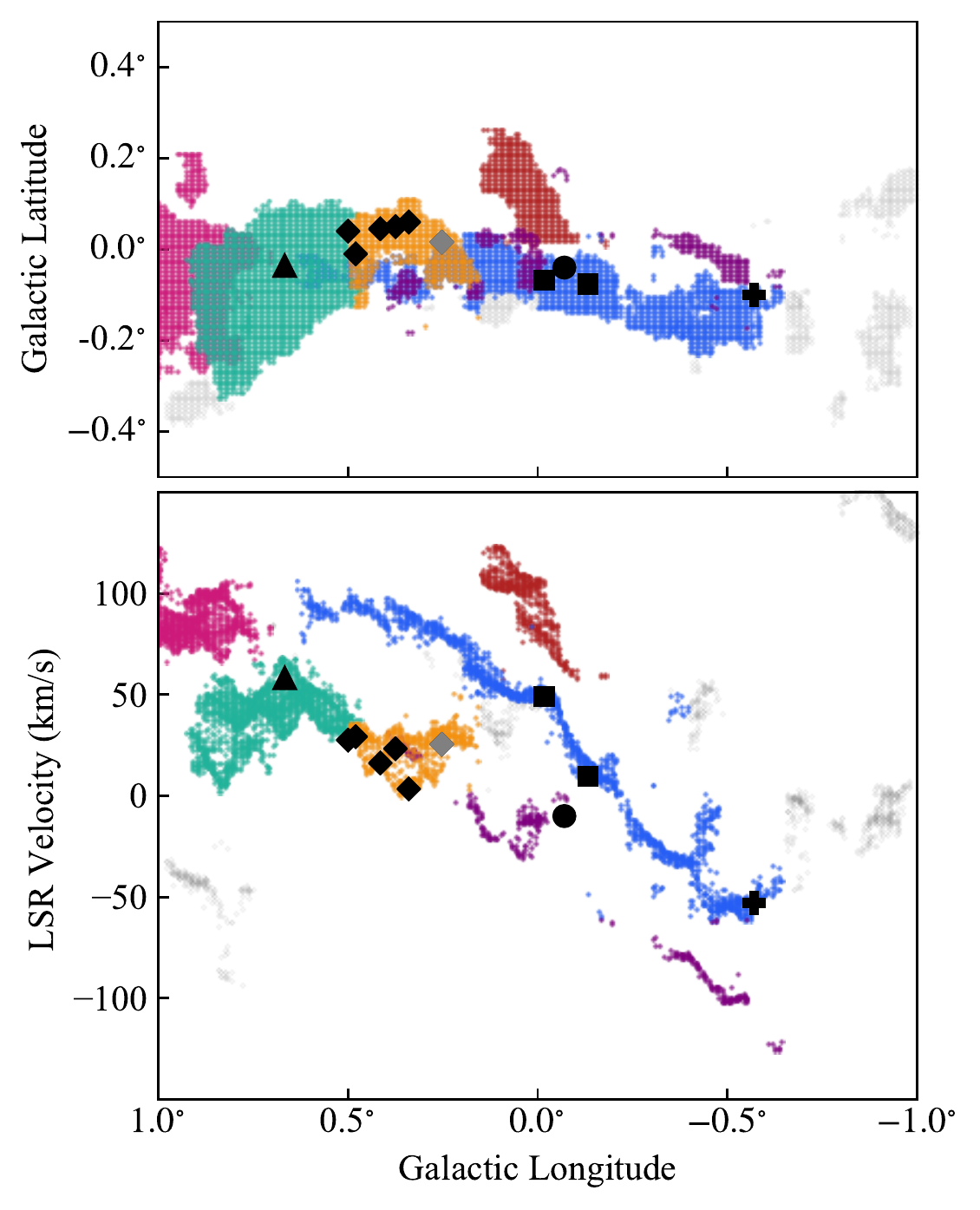}
\caption{($l,b$) (\textit{upper panel}) and ($l,v$) (\textit{bottom panel}) distribution of the NH$_\mathrm{3}$~(1,1) emission data from the HOPS survey fitted with \textsc{SCOUSE} \citep{Henshaw+16}. The Galactic longitude and latitude of each pixel denote the location of a spectral component with a corresponding best-fitting solution, while the LSR velocity denotes its centroid velocity. Different colours highlight some of the main structures in the ($l,b,v$) space: Arm I (purple),  Arm II (blue), Arm III (red), Sagittarius B2 complex (aqua), dust ridge (yellow), Complex $1.\!\!^{\circ}3$ (magenta). Different markers indicate the locations of prominent molecular clouds: Sagittarius B2 (triangle), dust ridge clouds (diamonds), Sagittarius C (cross), $20 \,\kms$ and $50\,\kms$ clouds (squares). The location of Sagittarius A* is denoted with a circle.}
\label{SCOUSE}
\end{figure}

In this Section, we summarise the main observed features of the CMZ, with the goal of familiarising readers with the structures to which we will compare our simulations below. We particularly focus on the region at longitude $\vert l \vert \leq 2^{\circ}$ and latitude $\vert b \vert \leq 0.\!\!^{\circ}5$ that hosts most of the dense gas emission. \autoref{NH3-CO} shows the intensity contours of CO~($J=1-0$) emission overlapping the distribution of NH$_\mathrm{3}$~(1,1) emission in the region of interest. The CO~($J=1-0$) emission data are taken from the CfA Galactic Plane survey \citep{Bitran+97, Dame+01}, while the NH$_\mathrm{3}$~(1,1) emission data are taken from the H$_\mathrm2$O Southern Galactic Plane Survey \citep[HOPS,][]{Walsh+11, Purcell+12}. The top plot shows the $(l,b)$ distribution integrated over the local standard of rest (LSR) velocity, $v$, while the bottom plot shows the longitude-velocity distribution, $(l,v)$, averaged over $b$. The overall emission distribution is similar for the two molecules: it spans $\sim 200 \kms$ in velocity and it is highly asymmetric, with more emission coming from positive longitudes and velocities. We remind readers that NH$_\mathrm{3}$~(1,1) traces denser gas than CO~($J=1-0$), and that the former line is usually optically thin, while the latter is certainly optically thick. This explains why the emission peaks of the two molecules do not necessarily coincide.

\autoref{SCOUSE} shows the NH$_\mathrm{3}$~(1,1) data from the HOPS survey fitted with the \textsc{SCOUSE} algorithm \citep{Henshaw+16}. Each point indicates the Galactic longitude, latitude and centroid line-of-sight velocity of a spectral component as determined by \textsc{SCOUSE}. Different colours highlight some of the main structures in the $(l,b,v)$ data cube, while different markers represent the location of the most prominent molecular clouds. In the following, we describe the main properties of these structures \citep[see][for a detailed description]{Henshaw+16, Longmore+17}.
\begin{itemize}
\item \textit{Arm I \& II.} Arm I (purple) and Arm II (blue) are two extended and almost parallel features in the longitude-velocity plot ($-0.\!\!^{\circ}65 \leq l \leq 0.\!\!^{\circ}5$, $-150 \kms \leq v \leq 100 \kms$). \citet{Sofue95} was the first to refer to these structures as ``Arms'', since in its interpretation they are two spiral arms within the CMZ.
\item \textit{Arm III.} Arm III (red), also known as ``polar arc'', is a structure located at high velocities and inclined by $> 40^{\circ}$ with respect to the Galactic plane.
\item \textit{Sagittarius B2.} Sgr B2 (aqua color and black triangle) is a prominent and highly star-forming molecular cloud complex ($0.\!\!^{\circ}5 \leq l \leq 0.\!\!^{\circ}85$, $-0.\!\!^{\circ}15 \leq b \leq 0.\!\!^{\circ}10$, $10 \kms \leq v \leq 70 \kms$). It presents a high number of independent velocity components at any given ($l,b$) and is characterised by a very broad range of velocity dispersions \citep[$13.3\kms<\sigma<53.1\kms$,][]{Henshaw+16}. Its complex kinematic structure can explain the shell-like features of Sgr B2, which are thought to be the result of a cloud-cloud collision \citep[e.g.][]{Hasegawa+94, Sato+00}. A recent alternative interpretation explains the complex structure of Sgr B2 as the superimposition of fragments belonging to the same cloud along the line of sight \citep{Henshaw+16, Kruijssen+19}.
\item \textit{Dust ridge.} Dust ridge (yellow) is the name used to indicate a sequence of molecular clouds (diamonds) located between Sgr2 and  G0.253+0.016 (grey diamond), also known as The Brick. The dust ridge clouds appear as absorption features at mid-infrared wavelengths, thus indicating the presence of cold and dense material \citep{Molinari+11}. \citet{Kruijssen+15} and \citet{Henshaw+16} identify two features parallel in velocity in the emission region associated to the dust ridge, suggesting that one of them is probably not associated with the dust ridge clouds.
\item \textit{$20\, \kms$ and $50\,\kms$ clouds.} The $20 \,\kms$ and $50\,\kms$ clouds (squares) are two bright molecular clouds coherently connected in ($l,b,v$) space \citep[e.g.][]{Bally+88, Coil&Ho00} located in projected proximity to Sgr A$^*$ (circle). A kinematic analysis of their emission shows that they are likely linked to Arm II \citep{Henshaw+16}. Similar to the dust ridge clouds, the $20 \kms$ and $50\kms$ clouds also appear as absorption features at mid-infrared wavelengths \citep{Molinari+11}. 
\item \textit{Sagittarius C.} Sgr C (cross) is a molecular cloud complex that is a site of active star formation \citep[e.g.][]{Yusef-Zadeh+09}. The bulk of its emission appears to be kinematically associated with the negative-latitude end of  Arm II \citep{Henshaw+16}.
\item \textit{Complex $1.\!\!^{\circ}3$.} The magenta dots denote the extension at low longitudes of Complex $1.\!\!^{\circ}3$, a huge molecular complex that extends over more than $1^{\circ}$ in longitude, suggested to be the main accretion site of material on to the CMZ \citep{Rodriguez+08}.
 \end{itemize}
 
\section{Methods}
\label{Methods}

\subsection{Numerical Scheme}
\label{Code}

We carry out our simulations using \textsc{GIZMO} \citep{Hopkins15}, a parallel magneto-hydrodynamical code, based on a mesh-free, Lagrangian finite-volume Godunov method. We use the Meshless Finite Mass solver on a reflecting-boundary domain and assume gas to follow an ideal equation of state with a constant adiabatic index $\gamma = 5/3$. 
We refer the reader to \citetalias{Armillotta+19} for a detailed description of the numerical methods used in our simulations. Here, we briefly summarize the relevant features:
\begin{itemize}
\item \textit{Gravity.} Self-gravity is included in our simulations and it is solved by the Tree method solver described in \citet{Springel05}. In addition to gas self-gravity, our simulations are performed in the presence of an external gravitational potential specific for our problem. We use the best-fit Milky Way potential by \citet{McMillan17} as modified by \citet{Ridley+17}, which includes contributions from the dark matter halo, the stellar disc, the bulge and the Galactic bar rotating with a constant patter speed $\Omega_\mathrm{p} = 40 \: \mathrm{km \, s^{-1} \, kpc^{-1}}$ \citep[e.g.][]{Wegg+15}. 
\item \textit{Radiative processes.} Radiative cooling is tracked through the astrophysical chemistry and cooling package \textsc{GRACKLE} \citep{Smith+17}, run in equilibrium mode. \textsc{GRACKLE} provides cooling rates for both primordial species and metals via look-up tables calculated through the \textsc{CLOUDY} spectral synthesis code \citep{Ferland+13} as a function of temperature and metallicity under the assumption of collisional ionisation equilibrium. In addition to radiative cooling, gas can also be heated via diffuse photoelectric heating in which electrons are ejected from dust grains via far-ultraviolet photons. This is implemented as a constant heating rate per hydrogen atom uniformly through-out the simulation box. We use a rate of of $8.5\times10^{-26}$~erg~s$^{−1}$, consistent with the expected heating rate for the inner regions of the Galaxy \citep[$R\sim3$~kpc,][]{Wolfire+03}.
\item \textit{Star formation.} A gas particle can be converted into a star particle if it is: 1) self-gravitating, i.e. virial parameter smaller than 1, 2) dense, i.e. gas density larger than $10^3$~cm$^{-3}$, 3) self-shielded \citep[according to the definition by][]{Krumholz&Gnedin11}, 4) not photoionised, i.e. gas temperature lower than $10^4$~K. If all these criteria are satisfied, we calculate the probability, $P$, for a given gas particle to be turned into a star as the fraction of gaseous mass that should be converted in stellar mass during the hydro time-step according to the local SFR. The SFR is parametrised as $\dot {\rho}_\mathrm{SF} = \epsilon_\mathrm{ff} {\rho_\mathrm{g}}/{t_\mathrm{ff}}$, where $\rho_\mathrm{g}$ is the local gas density, $t_\mathrm{ff} = \sqrt{{3 \pi}/{32 G \rho_\mathrm{g}}}$ is the local free-fall time and $\epsilon_\mathrm{ff} = 0.01 $ is the star formation efficiency \citep[this value is chosen based on observational evidence, e.g.][]{Krumholz&Tan07, Utomo+18}. If $P$ is larger than a randomly generated number $N \in [0,1)$, the gas particle is converted into a star particle with the same mass and dynamical properties. 
\item \textit{Stellar evolution.} We model the time evolution of each newly-formed star by using the stellar population synthesis code \textsc{SLUG} \citep{daSilva+12, Krumholz+15b}. Each star particle spawns an individual \textsc{SLUG} simulation that stochastically draws individual stars from a Chabrier initial mass funtion \citep{Chabrier05}, tracks their mass- and age-dependent ionising luminosities and determines if and when they explode as type II supernovae, along with mass, momentum and energy injection rates for each supernova. In the \textsc{SLUG} calculations we use the Padova stellar evolution tracks \citep{Bressan+12}, the ``starburst99'' spectral synthesis method of \citet{Leitherer+99} and Solar metallicity yields from \citet{Sukhbold+16}.
\item \textit{Stellar feedback.} Simulations include stellar feedback from supernova explosion and photoionisation, both of them implemented following the prescriptions in \citet{Hopkins+18b, Hopkins+18a}. Mass, momentum and energy injected by each supernova are distributed among the neighbouring gas particles in proportion to the solid angle centred on the star and subtended by the effective face between the gas particle itself and the star. We account for the unresolved energy-conserving phase of the blastwave by imposing a lower limit of $4 \times 10^5\,\mo\kms$ for the amount of momentum deposited in the ambient medium \citep[see e.g.][]{Kim&Ostriker15, Gentry+19}. The ionising luminosity from each star particle is instead distributed to neighbouring gas particles based on their distance from the star. Starting from the closest, each particle gains the amount of luminosity needed to be ionised, i.e. needed to reach a temperature of $10^4$~K, until the ionising photon budget is exhausted.
\end{itemize}

\subsection{Simulation overview}
\label{SimOverview}

The initial conditions of the simulation presented in \citetalias{Armillotta+19} consist of $2\times10^6$ gas particles with mass $2\times10^3\,\mo$ distributed in a cylindrical slab with radius 4.5~kpc and half-height 1~kpc and sampled in order to reproduce the mass distribution of the Galactic disc given by \citet{Binney&Tremaine08}. The field velocity is initialized so that gas particles are in radial equilibrium on circular orbits in the axysymmetric part of the gravitational potential of the Galaxy. In the first part of the simulation, gas evolves in the presence of pure hydrodynamics, external gravity and radiative cooling down to $10^4$~K. Once gas reaches a steady state equilibrium in the non-axisymmetric part of the gravitational potential, self-gravity, star formation, stellar feedback and full radiative cooling/heating are switched on.

\begin{figure*}
\includegraphics[width=\textwidth]{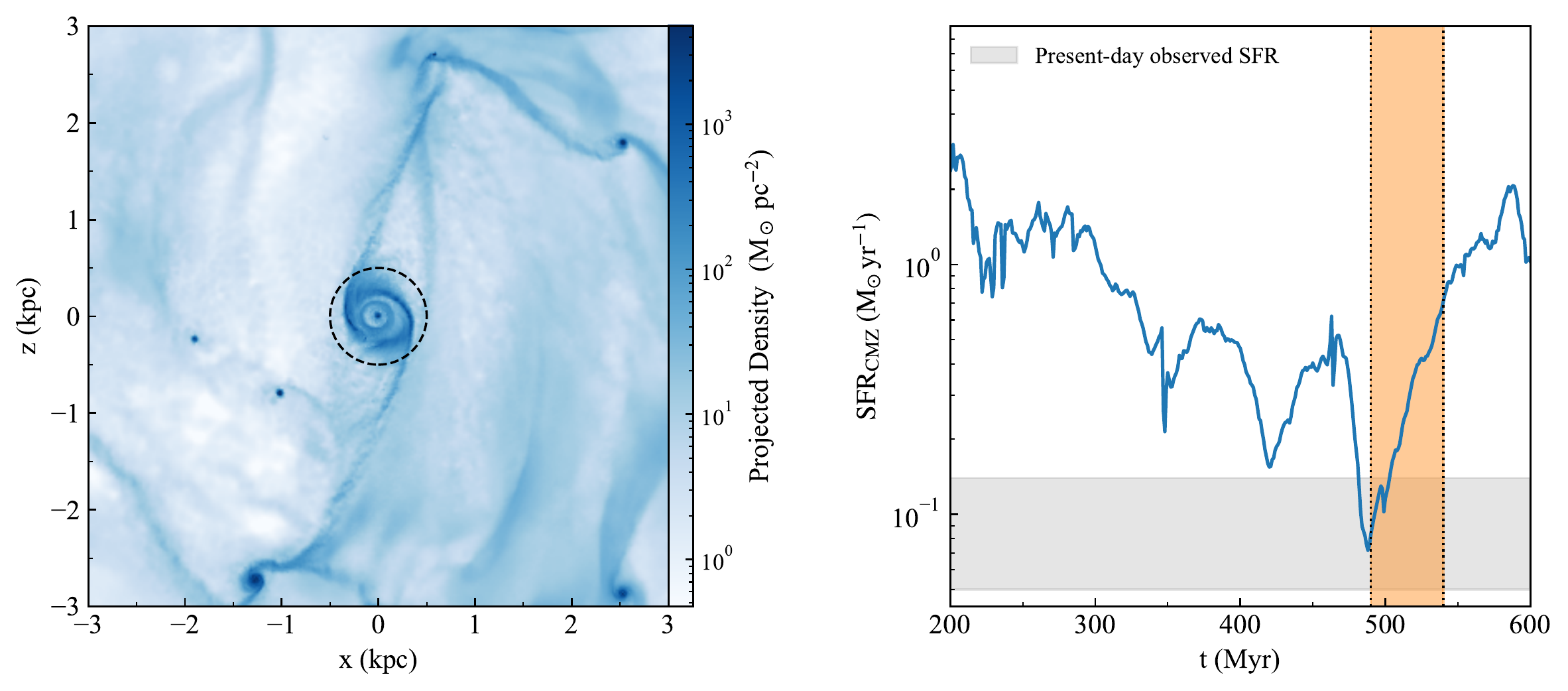}
\caption{Outcomes of the simulation presented in \citetalias{Armillotta+19}. \textit{Left panel}: Face-on gas density projection in the central 3~kpc region of the Galaxy. The black dotted circle encloses the region within 500~pc from the Galactic center, that we identify as the CMZ. The snapshot have been taken at $t=500$~Myr. \textit{Right panel}: Time evolution of the SFR in the CMZ. The grey bar indicates the approximate present-day observed SFR. The orange bar encompasses the temporal range between $t=490$~Myr and $t=540$~Myr of the high-resolution simulation analysed in this paper.}
\label{LowResSim}
\end{figure*}

The left panel of \autoref{LowResSim} shows an example of the face-on gas density projection in the central 3~kpc region of the Galaxy. In \autoref{Introduction}, we have briefly explained the gas dynamics in the central region of barred galaxies. Dissipative processes cause material in the outer parts of the bar to slowly flow inwards following $x_1$ orbits. When the $x_1$ orbits become self-intersecting, gas is shocked towards the Galactic centre, where it settles onto a ring-like structure at $R\sim 200-300$~pc and approximately follows $x_2$ orbits. The transition from the $x_1$ to the $x_2$ orbits happens through the so-called dust-lanes, visible in the left panel of \autoref{LowResSim} as dense features connecting gas orbiting at $R\sim3$~kpc to the central ring. The right panel of \autoref{LowResSim} shows the SFR evolution across time in the CMZ region. Burst/quenching cycles of star formation on time-scales of $\sim 50$~Myr are mainly driven by supernova feedback instabilities. A comparison with the present-day SFR of the CMZ ($\sim 0.04-0.15\,\moyr$, grey bar) suggests that it might lie at a minimum of a star formation cycle.

In this work, we re-simulate part of the simulation presented in \citetalias{Armillotta+19} at higher mass resolution, $200 \mo$, and lower gravitational softening length, 0.1~pc (i.e. a factor of 10 smaller mass and softening length than our previous simulation). We extract the initial conditions from the low-resolution run at $t = 490$~Myr, i.e. when the SFR is $\lesssim 0.1 \, \moyr$, and it is therefore representative of the present-day Milky-Way's CMZ, and run the simulation for 50 Myr. The temporal window of the high-resolution simulation is highlighted with an orange bar in the right panel of \autoref{LowResSim}. We choose our start time and run duration because the SFR increases by one order of magnitude in 50 Myr starting at 490 Myr, so that we will be able to characterise the gas distribution in two very different parts of the CMZ life cycle.

\subsection{Chemical and observational post-processing}
\label{Postprocess}

We use the high-resolution simulation of the CMZ to investigate the distribution of \hi, CO~($J=1-0$), NH$_\mathrm{3}$~(1,1) and HCN~($J=1-0$) emission both in ($x,y,z$) and ($l,b,v$) space. The \hi\ emission provides a view of the warmer atomic gas distribution within the CMZ, while the distributions of CO~($J=1-0$), NH$_\mathrm{3}$~(1,1) and HCN~($J=1-0$) emission map molecular gas regions at different densities. Among them, the CO~($J=1-0$) transition produces the brightest emission line. However, its relatively-low critical density ($\sim 1000$~cm$^{-3}$) and generally high optical depth ($\sim 10-100$) makes this transition inadequate to trace the high-density gas regions. NH$_\mathrm{3}$~(1,1) and HCN~($J=1-0$) transition lines are more suitable to trace gas at densities $\gg 10^3$~cm$^{-3}$. Of these two, the HCN line is generally brighter due to its higher Einstein $A$ coefficient, but by the same token is more vulnerable to optical depth effects.

Since we do not follow the gas chemical evolution in our simulation, we infer the luminosity associated to a given emission line using the post-processing astrochemistry and radiative transfer code \textsc{DESPOTIC} \citep{Krumholz14}. 
\textsc{DESPOTIC} calculates the chemical and thermal state of an optically-thick spherical cloud given its physical properties (e.g. density, non-thermal velocity dispersion), its chemical composition, the abundances of the elements within it and the radiation field around it. The chemical equilibrium calculation uses the C-O chemical network of \citet{Gong+17}, while the thermal equilibrium calculation includes heating by cosmic rays, grain photoelectric effect, cooling by the lines of \hi, \cii, \ci, \oi\ and CO, and collisional energy exchange between dust and gas. We use \textsc{DESPOTIC}'s large velocity gradient (LVG) option to calculate optical depths for line cooling, and we adopt Solar abundances for all elements in the chemical network. The nitrogen-bearing species are not included in \citeauthor{Gong+17} network, so we assume that these species are present in any gas that is sufficiently shielded for CO to be the main repository of carbon, with the same ratio to CO as is generally found in local observations. Specifically, we adopt $X_{\rm HCN} = X_{\rm NH_3} = 10^{-8} \left(X_{\rm CO}/X_{\rm C}\right)$, where $X_{\rm S}$ indicates the number of members of species S per H nucleus, and $X_{\rm C}=1.1\times 10^{-4}$ is the abundance of all carbon atoms regardless of chemical state. Thus in gas where all carbon is in the form of CO, the abundances of HCN and NH$_3$ are $10^{-8}$; see \citet{Offner+08} and \citet{Onus+18} for further discussion of these choices. We refer readers to \citet{Krumholz14} for details on DESPOTIC in general. 

We use \textsc{DESPOTIC} to generate four tables where \hi\ and H$_\mathrm{2}$ mass fractions, $X_\mathrm{HI}$ and $X_\mathrm{H_2}$, CO~($J=1-0$), NH$_\mathrm{3}$~(1,1) and HCN~($J=1-0$) transition line luminosities, $L_\mathrm{CO}$, $L_\mathrm{NH_3}$ and $L_\mathrm{HCN}$, and gas temperature, $T_\mathrm{g}$, are listed as a function of hydrogen number density, $n_\mathrm{H}$, hydrogen column density, $N_\mathrm{H}$, and velocity gradient, $dv/dr$. Each table is produced assuming a different set of values for the strength of the interstellar radiation field (ISRF), $\chi$, and the primary cosmic ray ionization rate, $\zeta$. We investigate four cases: \textit{Solar} radiation field, $\chi = 1\,G_0$\footnote{$G_0$ corresponds to the ISRF strength given in \citet{Draine78}.} and $\zeta = 10^{-16}$~s$^{-1}$; \textit{weak} CMZ radiation field, $\chi = 10^2 \,G_0$ and $\zeta = 10^{-15}$~s$^{-1}$; \textit{intermediate} CMZ radiation field, $\chi = 10^{2.5} \,G_0$ and $\zeta = 10^{-14.5}$~s$^{-1}$; \textit{strong} CMZ radiation field, $\chi = 10^3 \,G_0$ and $\zeta = 10^{-14}$~s$^{-1}$. The \textit{Solar} case is representative of the solar neighbourhood \citep[e.g.][]{vanderTak+00, LePetit+04, Indriolo+McCall}. However, the ISRF and the cosmic-ray fluxes in the inner regions of the Galaxy are expected to be significantly higher than those measured in molecular clouds in the local disk \citep[e.g.][]{Wolfire+03}; thus the \textit{Solar} case is provided solely as a baseline for comparison. Unfortunately, ionisation rates and ISRF strengths in the CMZ region are poorly constrained, and are in reality probably not uniform in space or time. Published estimates span at least a factor of 10, depending on the technique used and the region being observed. Our range from \textit{weak} to \textit{strong} cases is intended to be roughly representative of the range of published estimates \citep[e.g.][]{Clark+13, Ginsburg+16, Oka+19}. We discuss the choice of radiation field model further in \aref{append}, where we show that our \textit{intermediate} case provides the best overall fit to the present-day CMZ, and that it successfully reproduces bulk properties such as the total molecular gas mass and CO luminosity. We therefore adopt this choice for all the results presented in the main text, deferring discussion of how the results depend on the radiation field to the Appendix.

We post-process the simulation outcomes to calculate $X_\mathrm{HI}$, $X_\mathrm{H_2}$, $T_\mathrm{g}$, $L_\mathrm{CO}$, $L_\mathrm{NH_3}$ and $L_\mathrm{HCN}$ as follows. We first compute the values of $n_\mathrm{H}$, $N_\mathrm{H}$, and $dv/dr$ for each gas particle and then perform a trilinear interpolation on the \textsc{DESPOTIC} tabulated values. The hydrogen number density is calculated as 
\begin{equation}
n_\mathrm{H} = \dfrac{\rho \mu_\mathrm{H}}{m_\mathrm{H}}\,,
\end{equation}
where $\rho$ is the local gas density associated to the gas particle, $\mu_\mathrm{H} = 0.71$ is the hydrogen mass fraction assuming solar metallicity and $m_\mathrm{H} = 1.67 \times 10^{-24}$~g is the mass of the hydrogen nucleus. The hydrogen column density is defined as in \citet{Safranek-Shrader+17},
\begin{equation}
N_\mathrm{H} = n_\mathrm{H} L_\mathrm{shield}\,,
\label{NH}
\end{equation}
where $L_\mathrm{shield}$ is the shielding length. Post-processing of galactic-disc-scale simulations with ray-tracing-based radiative transfer and chemical network integration have shown that local models with photo-shielding based on \autoref{NH} reproduces the distribution and amount of molecular gas reasonably well as compared with a detailed, global ray-tracing calculation. Specifically, an approach based on the Jeans length ($L_\mathrm{shield} = L_\mathrm{J} = \pi c_\mathrm{s}^2 / G \rho$, where $c_\mathrm{s}$ is the local adiabatic sound speed) with a $T = 40$~K temperature cap yields the most accurate H$_2$ and CO abundances \citep{Safranek-Shrader+17}. Finally, the velocity gradient is calculated as
\begin{equation}
\dfrac{dv}{dr}= \| \nabla \otimes {\bf{v}} \|  \equiv \sqrt{\sum_{i,j=1}^3\left(\dfrac{\partial v_i}{\partial x_j}\right)^2} \,,
\end{equation}
where ${\bf{v}}$ the local gas velocity associated to the gas particle. 

\begin{figure*}
\includegraphics[width=\textwidth]{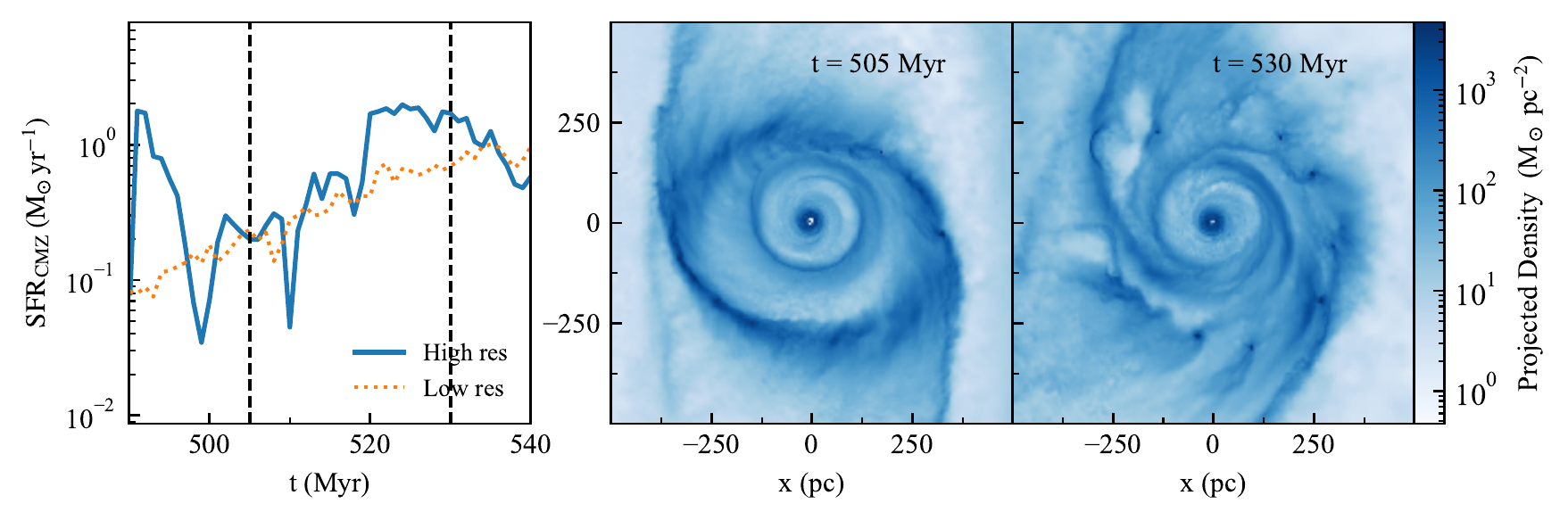}
\caption{CMZ region in the high-resolution simulation discussed in this paper. Time evolution of the SFR (blue solid line, \textit{left panel}) and face-on gas density projection at two different times, $t=505$~Myr (\textit{middle panel}) and $t=530$~Myr (\textit{right panel}). The orange line shows the time evolution of the SFR in the low-resolution simulation (see \autoref{LowResSim}). The two dashed lines in the SFR plot indicate the SFR values at the times when the two snapshots have been taken.}
\label{HighResSim}
\end{figure*}

Once the chemical post-process of the simulation is computed, we calculate the ($l,b,v$) gas particle distributions for each simulation snapshot. We assume that the Sun is undergoing circular motion in the Galactic plane at a radius $R_\odot = 8$~kpc with speed $v_\odot = 240 \kms$ and that the angle between the bar major axis and the Sun-Galactic Centre line is $20^{\circ}$ \citep[e.g.][]{Bland-Hawthorn+16}. We bin the gas particles within 500~pc from the Galactic Center in ($l,b,v$) space. The resolution of each bin is $\Delta l = 0.\!\!^{\circ}015$, $\Delta b = 0.\!\!^{\circ}015$ and $\Delta v = 1 \kms$. In each bin, we sum the brightness temperatures associated to the CO~($J=1-0$), NH$_\mathrm{3}$~(1,1) and HCN~($J=1-0$) emissions, $T_\mathrm{B,CO}$, $T_\mathrm{B,NH_3}$ and $T_\mathrm{B, HCN}$ respectively, and the \hi\ column density, $N_\mathrm{\hi}$, over all the gas particles within the bin. The brightness temperature is calculated as 
\begin{equation}
 T_\mathrm{B}[\mathrm{K}] =\dfrac{\lambda^2}{2 k_\mathrm{B}}\dfrac{L}{4 \pi D^2 \Delta \nu} \dfrac{180^2}{\pi^2{\Delta l \Delta b}} \,,
\end{equation}
where $\lambda$ is the wavelength associated to the line transition, which is 0.26~cm for the CO~($J=1-0$) transition, 1.27~cm for the NH$_\mathrm{3}$~(1,1) transition and 0.34~cm for the HCN~($J=1-0$) transition), $k_\mathrm{B} = 1.38 \times 10^{-16}$~erg~K$^{-1}$ is the Boltzman constant, $L$ is the emission line luminosity in erg/s associated to each gas particle, $D$ is the distance of the gas particle from the Sun in cm and $\Delta \nu = 10^5 \Delta v/ \lambda$ is the frequency resolution. The \hi\ column density of each particle is calculated as
\begin{equation}
N_\mathrm{HI} = n_\mathrm{HI}\,h = X_\mathrm{HI} n_\mathrm{H} \,h \,,
\end{equation}
where $h\equiv(M/\rho)^{-1/3}$ is the spatial resolution scale (average inter-particle separation around the gas particle), with $M$ the gas particle mass. We calculate the \hi\ column density, rather than the luminosity associated to the \hi-21~cm transition, since this is more informative. However, converting the \hi\ column density into luminosity is straightforward since \hi\ is mostly optically thin \citep{Dickey&Lockman90}. 

\section{Simulation analysis} 
\label{Results}

Before analysing the post-processed gas distribution, we give an overview of the CMZ evolution throughout the high-resolution simulation. Note that in the simulation we identify the central 500~pc region as CMZ. The left panel of \autoref{HighResSim} shows the SFR versus time (blue line). The simulation experiences a brief burst of star formation as soon as we increase the mass resolution / decrease the softening length. However, after a few Myr stellar feedback pushes the system back towards equilibrium, and the star formation rate drops to values comparable to those found at matching times in the lower-resolution simulation. From that point on, the SFR increases as a function of time, going from $\sim 0.1-0.2 \,\moyr$ to $1-2 \, \moyr$ in $20-30$ Myr. For comparison, we show the SFR as a function of time in the low-resolution simulation (dotted orange line). The time-averaged trends are very similar at the two different resolutions, but in the high-resolution case the SFR fluctuates more rapidly since we can capture deeper collapses followed by more effective episodes of feedback from the resulting highly-clustered stars. We refer readers to Appendix A of \citetalias{Armillotta+19} for a more thorough discussion on the resolution-dependence of our results.

In the right panels of \autoref{HighResSim}, we show the CMZ gas density distribution in two different moments of the star formation activity. When the level of star formation is low (SFR~$\sim 0.2 \, \moyr$ at $t=505$~Myr), most of the gas lies in a highly dense ($\Sigma_\mathrm{gas} \gtrsim 10 ^3 \,\mo$~pc$^{-2}$) elliptical ring located at $\sim 200-300$~pc from the Galactic center. In addition to the main ring, we can note the presence of a less dense ring/spiral-like structure located at $\sim 100$~pc from the Galactic center. We identify this gas with the wake of a massive molecular cloud ($M\geq10^7\,\mo$) tidally destroyed by its encounter with the Galactic centre (see Sec.~3.2 in \citetalias{Armillotta+19}) in the low-resolution run. The gas configuration is completely different when the CMZ is actively star-forming (SFR~$\sim 2 \, \moyr$ at $t=530$~Myr). Supernova feedback associated to the intense star formation activity enhances the level of turbulence, thus disrupting the dense ring. The new gas configuration consists in a number of molecular clouds rotating nearly in the same position of the former ring and surrounded by more diffuse turbulent gas. 

In the following, we focus on the analysis of the post-processed simulation in the two different moments of the CMZ life, close to the minimum (\autoref{LowSF}) and the maximum  (\autoref{HighSF}) of the star formation activity, and on the dense gas evolution across time (\autoref{DenseGas}). 

\subsection{Low SFR: synthetic maps}
\label{LowSF}

\begin{figure*}
\includegraphics[width=\textwidth]{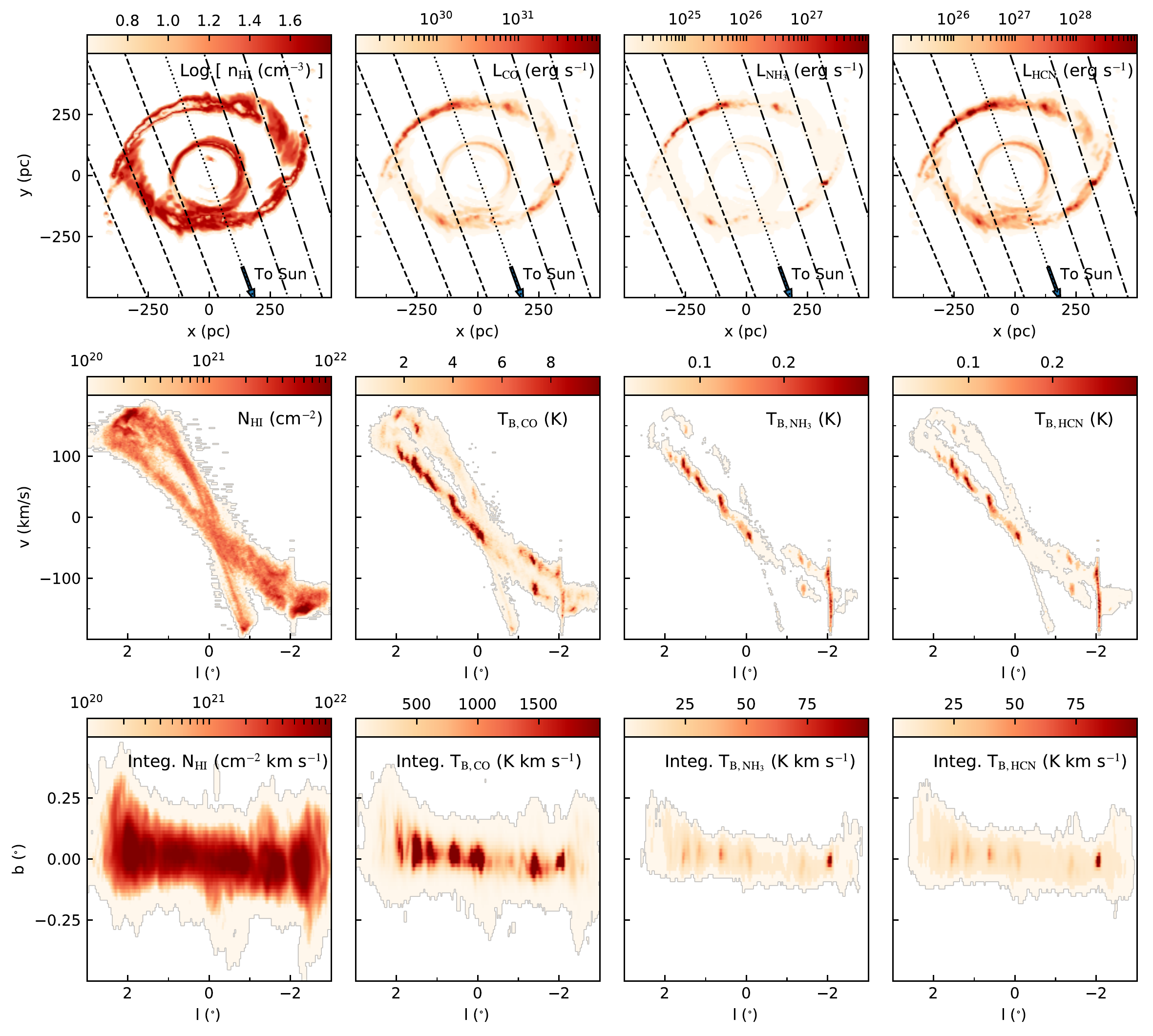}
\caption{\textit{Top panels}: from left to right snapshots taken at $z=0$ of \hi\ number density, $n_\mathrm{HI}$, CO~($J=1-0$) luminosity, $L_\mathrm{CO}$, NH$_\mathrm{3}$~(1,1)  luminosity, $L_\mathrm{NH_3}$, and HCN~($J=1-0$) luminosity, $L_\mathrm{HCN}$. The bar major axis is aligned with the axis $x = 0$. In each plot, the dotted black line indicates the line connecting the Sun to the Galactic Center, $l=0$, while the dashed/dot-dashed line indicate the line of sights at $l = \pm 1^{\circ},2^{\circ},3^{\circ}$. The arrow points towards the Sun position. \textit{Central panels}: from left to right longitude-velocity distributions of  $\hi$ column density, $N_\mathrm{HI}$, CO~($J=1-0$) brightness temperature, $T_\mathrm{CO}$, NH$_\mathrm{3}$~(1,1) brightness temperature, $T_\mathrm{NH_3}$, and HCN~($J=1-0$) brightness temperature, $T_\mathrm{HCN}$, averaged over the latitude range $\pm 0.\!\!^{\circ}5$. \textit{Bottom panels}: from left to right ($l,b$) maps of $N_\mathrm{HI}$, $T_\mathrm{CO}$, $T_\mathrm{NH_3}$, and $T_\mathrm{HCN}$, integrated over the velocity range $\pm 200 \, \kms$. The gas distribution is analysed at $t=505$~Myr.}
\label{Maps_LowSFR}
\end{figure*}

\begin{figure}
\includegraphics[width=0.49\textwidth]{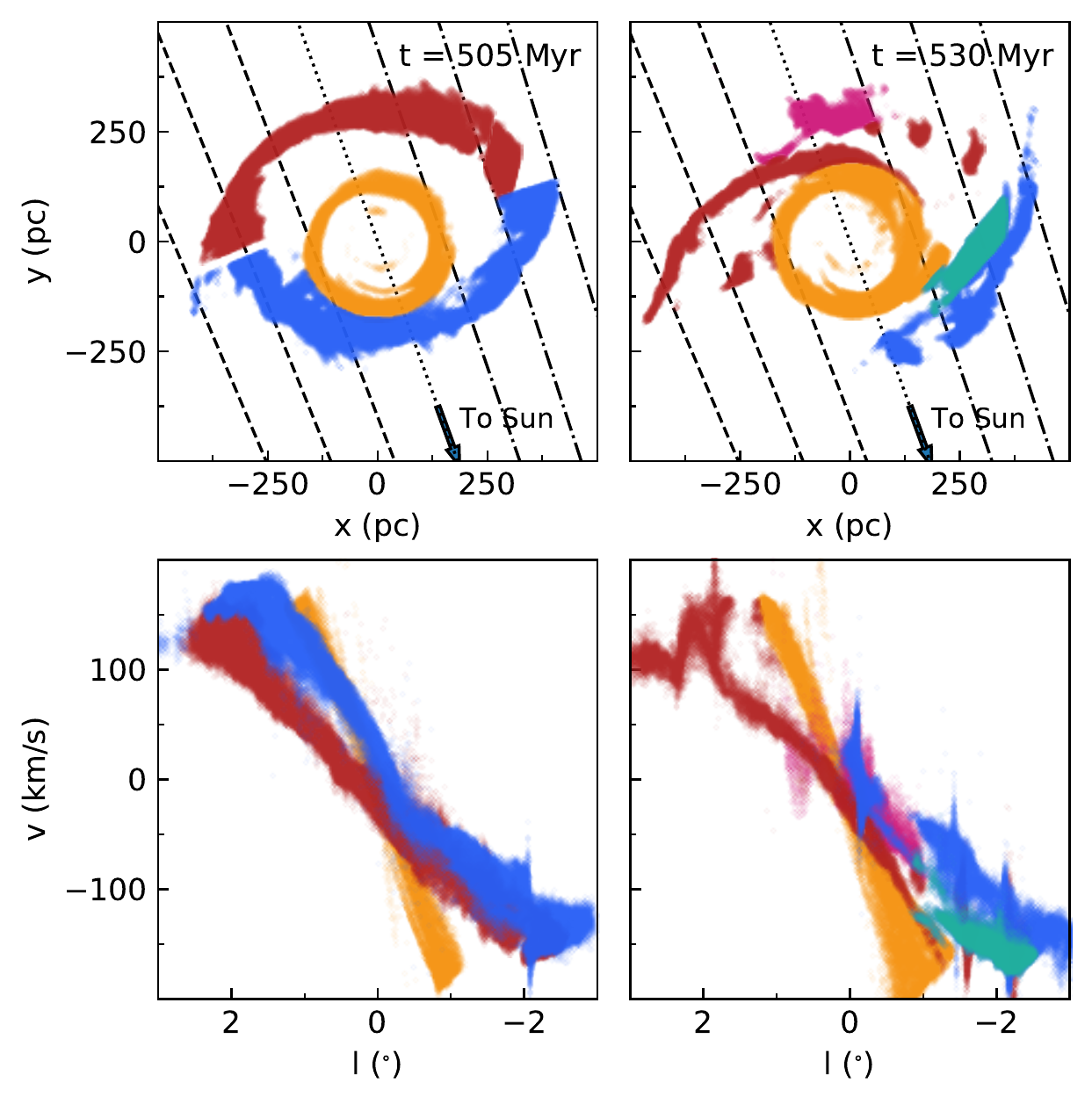}
\caption{Gas particle distribution colour coded to identify corresponding structures in the ($x,y$) and ($l,v$) space at $t=505$~Myr (\textit{left panels}) and $t=530$~Myr (\textit{right panels}). Only gas particles identified by \textsc{DESPOTIC} as atomic and molecular gas are shown.}
\label{Mapping}
\end{figure}

\begin{figure*}
\includegraphics[width=\textwidth]{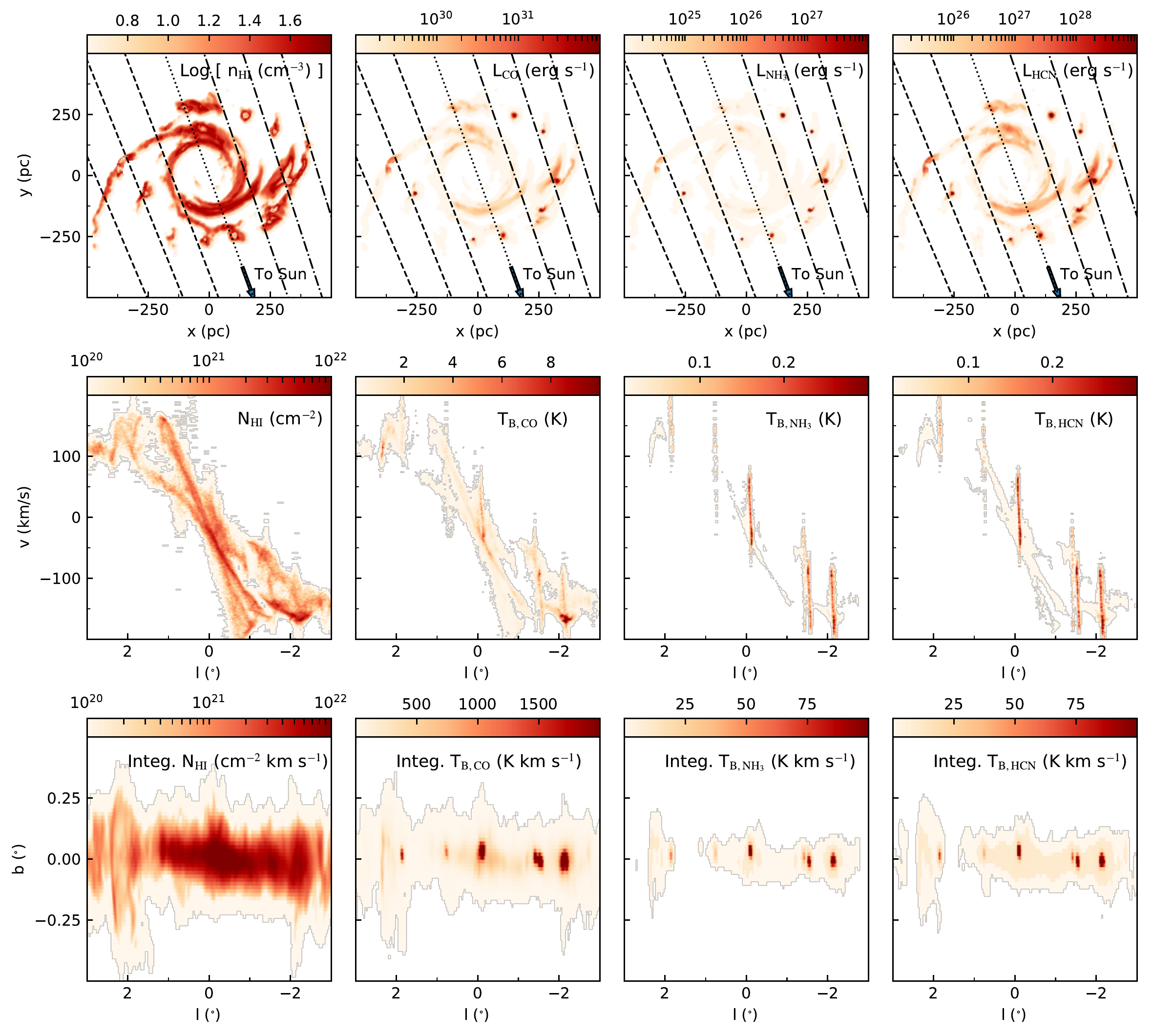}
\caption{Same as \autoref{Maps_LowSFR}, but at $t = 530$~Myr.}
\label{Maps_HighSFR}
\end{figure*}

In \autoref{Maps_LowSFR}, we present a view of the CMZ at $t=505$~Myr as seen through different gas tracers. The top panels show the Cartesian distribution of \hi\ number density and CO~($J=1-0$), NH$_\mathrm{3}$~(1,1) and HCN~($J=1-0$) luminosity taken at $z=0$. The central and bottom panels show the distribution of $N_\mathrm{HI}$, $T_\mathrm{CO}$, $T_\mathrm{NH_3}$, and $T_\mathrm{HCN}$ in the ($l,v$) and ($l,b$) diagram, respectively. We decide to not plot the gas distribution in the central few pc of the Galaxy. This region is unresolved in our simulations and, since we do not include the gravitational potential of the black hole that dominates the central few pc, the dynamics within it is not reliable. The distributions in the ($l,v$) diagram have been averaged over $ b = \pm 0.\!\!^{\circ}5$, while the distributions in the ($l,b$) diagram have been integrated over $v = \pm 200\,\kms$. The maps are not colour-coded in regions where the gas temperature calculated by \textsc{DESPOTIC} is above $2\times10^4$~K, i.e. when gas is fully ionised. \textsc{DESPOTIC} is indeed not intended to work at such temperatures. We observe that the dense structures in the ($l,v$) diagram show low velocity dispersion compared to the observed data (see \autoref{NH3-CO}). Indeed, due to the finite simulation resolution, we cannot trace the velocity distribution on scales smaller than the spatial resolution scale, thus the line-of-sight velocity component associated to each gas particle shows a delta-function profile rather than a Gaussian-like profile.  

The first thing to note from \autoref{Maps_LowSFR} is that the molecular ring formed in our simulation extends from $l = 2.\!\!^{\circ}5$ to $l = -2.\!\!^{\circ}5$. In \autoref{Introduction} and \autoref{Data}, we have seen that most of the molecular gas in the CMZ is observed within an angular distance of $1^{\circ}$ from the Galactic center. Thus, the size of the dense ring in our simulation is more than a factor of 2 larger. This is because the gravitational potential used in our simulation was built to be consistent with large scale properties of the Milky Way, but it was not fine-tuned to reproduce the observed extension of the CMZ. Recently \citet{Sormani+19} have modified the bar parameters in \citeauthor{Ridley+17} gravitational potential to match the observed size of the CMZ. They have performed simulations similar to ours, but without self-gravity and stellar feedback, and found that the CMZ gas configuration produced in response to the new gravitational potential is a smaller ring structure in better agreement with the observed CMZ size. However, the small inconsistency between our adopted global potential and the observed ring size should have relatively small effects on the dense gas, whose properties are more sensitive to self-gravity and feedback than to the overall potential.

In the top panels of  \autoref{Maps_LowSFR}, we can observe how the emission produced by each gas tracer is distributed within the CMZ ring. Among the molecular gas tracers, the CO~($J=1-0$) transition produces the brightest emission throughout the entire ring. However, given its low critical density, this transition is inadequate to map the high-density gas regions. For example, the bright filament in the upper side of the ring appears smooth and uniform in the $L_\mathrm{CO}$ map, while its clumpy internal structure is well resolved in the $L_\mathrm{NH_3}$ and $L_\mathrm{HCN}$ maps. The emission associated to NH$_\mathrm{3}$~(1,1) and HCN~($J=1-0$) is indeed able to map denser and isolated structures within the ring. The \hi\ distribution traces the diffuse atomic gas. Its density peaks in the gas envelope surrounding molecular clouds.

In the longitude-velocity diagrams, we identify the presence of two nearly-parallel elongated features extending from positive longitudes and velocities to negative longitudes and velocities. Their emission is more or less continuous depending on the gas tracer. We use the left panel of \autoref{Mapping} to identify corresponding structures in ($x,y$) and ($l,v$) space. The top feature (colour-coded in blue) corresponds to the ring side closer to the Sun, while the bottom feature (color-coded in red) corresponds to the ring side farther from the Sun.  Gas rotates clockwise, moving from positive longitudes and velocities to negative longitudes and velocities in the near ring side, and from negative longitudes and velocities to positive longitudes and velocities in the far ring side. Although the velocity field of $x_2$ orbits is symmetric with respect to the bar axes, the two sides of the ring appear as two offset features in the ($l,v$) plane because the orbit is elliptical and the Sun-Galactic Centre line is not aligned with one of the two axes of the ellipse. As a consequence, the line of sight crosses the two sides of the orbit at points that are not symmetric with respect to the axes of symmetry of the ellipse\footnote{Note that this statement should in principle be valid even though the Sun-Galactic Centre line were aligned with one of the two bar axes. Since individual lines of sight diverge from the Sun-Galactic Centre line with increasing distance from the Sun, they should intercept non-axysimmetric points on the near and far sides of elliptical orbits. However, the CMZ extension is so small compared to the Sun-Galactic Centre distance that the lines of sight intercepting the CMZ can be considered almost parallel, and thus this effect is small. The majority of the splitting in the ($l,v$) plane therefore is due to the offset between the ellipse axis and the sightline, rather than due to the divergence of sightlines.}, which results in a different projected rotational velocity for gas on the near and far sides. 

The \hi\ distribution is nearly uniform, thus allowing us to identify the entire ring structure in the longitude-velocity map. The emission appears less uniform in the CO map and it is predominantly clumpy in the NH$_3$ and HCN diagrams. The non-uniform molecular gas distribution is the consequence of non-uniform star formation, which leads to different local conditions of turbulence, radiative cooling and gravity within the ring. Asymmetries in the CMZ were also found in the simulations with no self-gravity and stellar feedback performed by \citet{Sormani+17}. However, in such simulations asymmetries develop due to the combination of thermal instability and the so-called ``wiggle instability''. We can observe that the emission associated to the bottom stream is brighter compared to the top stream, especially at positive longitudes and velocities, corresponding to the ring region connected to the dust lanes (see \autoref{DenseGas} for a more accurate analysis of these bright high-density structures). 
Finally, the low-density inner ring at $R\sim 100$~pc (color-coded in yellow in \autoref{Mapping}) appears as a less inclined feature in the \hi\ ($l,v$) map, but its emission becomes weaker and weaker moving towards dense gas tracers. We point out that the match in extension between the inner ring formed in the simulation ($\vert l \vert \leq 1^{\circ}$) and the observed CMZ is just a coincidence and no connection can be traced between these two structures. The observed CMZ is indeed highly dense and star-forming, similarly to the external ring formed in the simulation.

In the longitude-latitude diagrams, we observe that the ring is slightly tilted with respect to the Galactic plane, especially on the positive longitude side. The NH$_3$ and HCN emission peaks, corresponding to the densest structures in the CMZ, are located within $\pm 0.\!\!^{\circ}05$ from the plane.

\subsection{High SFR: synthetic maps}
\label{HighSF}

\autoref{Maps_HighSFR} is the same as \autoref{Maps_LowSFR}, but at $t = 530$~Myr. As anticipated in \autoref{HighResSim}, the gas distribution is much less uniform when the CMZ is actively star-forming because feedback associated with the star formation activity shreds the dense ring. The lack of a coherent structure is clearly visible also in the more diffuse atomic gas phase. In the \hi\ ($l,v$) plot, the two parallel and elongated features present at $t=505$~Myr are replaced by multiple ridges of gas with lower extension in longitude and velocity, corresponding to fragments of the previous ring, or by arc-like structures (e.g. structure at $v \sim 100-150 \, \kms$ and $l\sim 2^{\circ}$), corresponding to gas compressed by supernova explosions. The low-density inner ring is the only structure that is much the same in \autoref{Maps_LowSFR} as in \autoref{Maps_HighSFR}, since no star formation occurs in it. 

Strong NH$_3$ and HCN emission is produced by dense and star-forming molecular clouds. In the ($l,v$) plot, these clouds appear like features with narrow longitude range and broad velocity dispersion ($\sigma \lesssim 50 \, \kms$). Although high velocity dispersions are in fact measured in the Milky-Way CMZ (e.g. Sgr B2, see \autoref{Data}), we stress that the internal dynamics of the high-density clouds produced in our simulation are unreliable. As already discussed in \citetalias{Armillotta+19}, gas at sufficiently high density is not resolved due to the presence of softened gravity, which prevents structures smaller than the gravitational softening length from forming. In the simulation analysed in this paper the softening length is 0.1~pc. In a Lagrangian code, like \textsc{GIZMO}, the inter-particle separation is directly related to the gas density, $h\equiv(M/\rho)^{-1/3}$. Thus, given our mass resolution of $200\,\mo$, the distance between gas particles approaches the gravitational smoothing length when $n_\mathrm{H} \gtrsim 10^6$~cm$^{-3}$. We therefore underestimate the velocity dispersion of gas approaching this density.

Finally, in the right panels of \autoref{Mapping}, we note that, despite the much less coherent gas distribution, gas located on the near side with respect to the Sun is still characterized by line-of-sight velocities higher than those of gas located on the far side at equal longitude. This is because the centres of mass of individual gas clumps nearly rotate on elliptical $x_2$ orbits \citep[see also][for similar results]{Sormani+17}.

\subsection{High-density gas evolution}
\label{DenseGas}

\begin{figure*}
\includegraphics[width=0.87\textwidth]{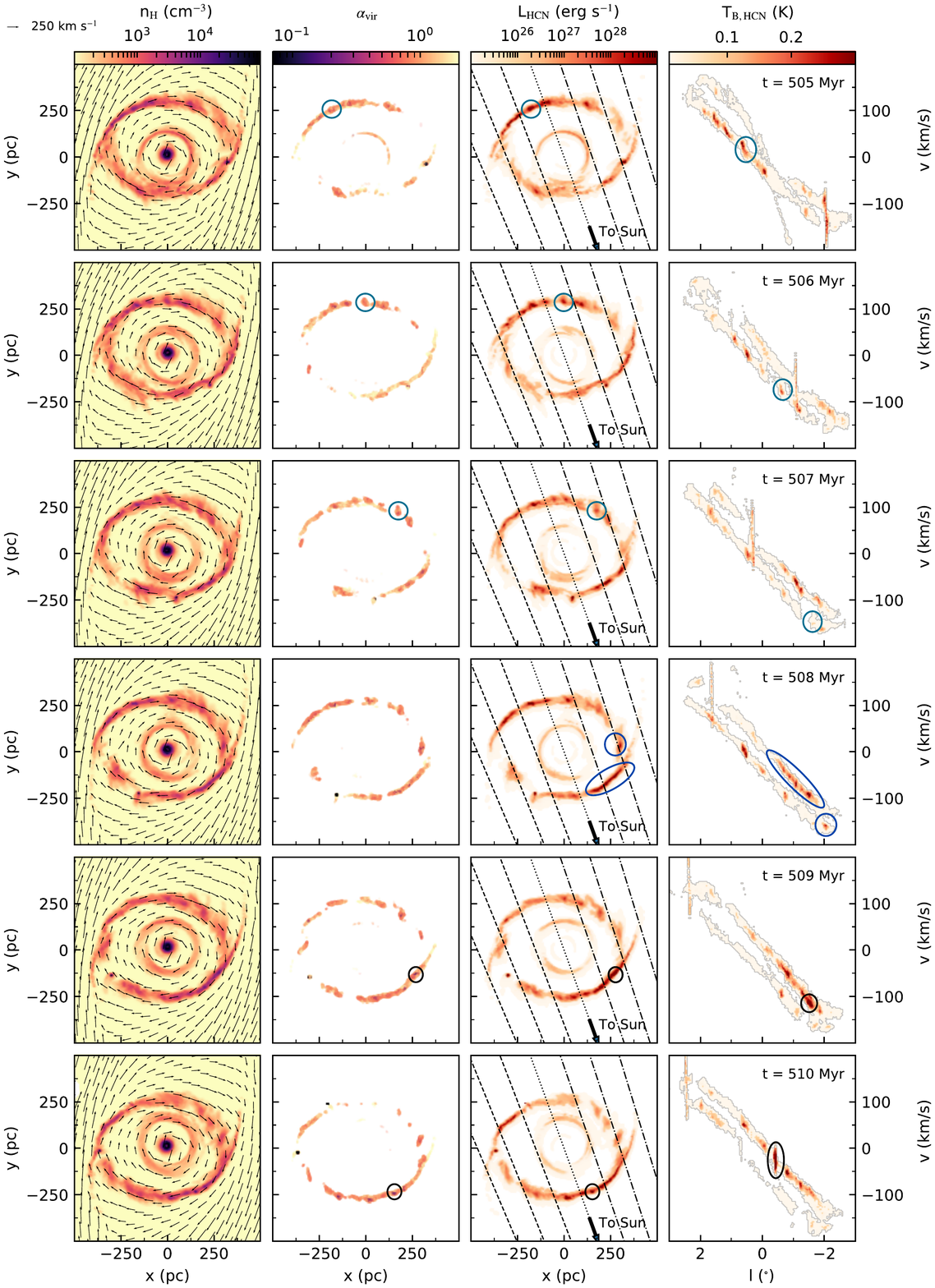}
\caption{Temporal sequence of snapshots taken at $z=0$. Each row shows the Cartesian distribution of $n_\mathrm{H}$ (\textit{first column}), virial parameter (\textit{second column}), $L_\mathrm{HCN}$ (\textit{third column}) and the ($l,v$) distribution of $T_\mathrm{B,HCN}$ (\textit{fourth column}) at a given time. The time at which the snapshots have been taken is indicated in the last panel of each row. In the first panel of each row the velocity field overlaps the $n_\mathrm{H}$ distribution. The velocity field is shown with vectors, whose length indicates the velocity magnitude. Coloured contours denote relevant structures discussed in the main text.}
\label{CMZEvolution}
\end{figure*}

In this Section, we focus on the high-density gas ($n_\mathrm{H} \gtrsim 10^3$~cm$^{-3}$) evolution and its distribution in longitude-velocity space. \autoref{CMZEvolution} shows the time evolution of the CMZ region seen through different indicators. The first panel of each row shows the velocity field overlapping the hydrogen gas number density distribution in the frame co-rotating with the bar. The second panel shows the virial parameter, $\alpha_\mathrm{vir}$, distribution for gas at $n_\mathrm{H} \geq 10^3$~cm$^{-3}$, which gives an indication of the gravitational stability of the gas. The virial parameter is defined as 
\begin{equation}
\alpha_\mathrm{vir} = \dfrac{(dv/dr)^2 + (c_\mathrm{s}/h)^2}{8 \pi G \rho}\,.
\end{equation}
The third panel shows the HCN~($J=1-0$) luminosity distribution, while the fourth panel displays the HCN~($J=1-0$) brightness temperature ($l,v$) map. We investigate the temporal range between 505 and 510~Myr, i.e. when the SFR is roughly consistent with the present-day value. 

As anticipated in \autoref{LowSF}, the longitude-velocity HCN distribution is highly asymmetric, and most of the emission comes from the ring regions immediately connected to the dust lanes (see \autoref{Maps_LowSFR} and \autoref{Mapping}). A comparison between the $n_\mathrm{H}$ and $L_\mathrm{HCN}$ maps shows that this region is actually very dense ($n_\mathrm{H}  \gg 10^3$~cm$^{-3}$). In this region gas flowing towards the Galactic Center through the dust lanes collides with gas orbiting within the ring, thus enhancing the local gas density. An example is provided in the panels at $t=508$~Myr and $t=509$~Myr. We highlight two regions in the  $L_\mathrm{HCN}$ and $T_\mathrm{B,HCN}$ maps at $t=508$~Myr: the region connected to the dust lane (blue ellipse) in the ring side near the Sun and a dense cloud (blue circle) which is crushing into it. This collision results in an increase of the gas density and emission at $t=509$~Myr.
\autoref{CMZEvolution} shows however that features such as this are not necessarily present at other times. For example, at $t=505$~Myr the ($l,v$) plot is characterised by a bright ridge of clouds at positive longitudes associated to the ring side far from the Sun. This feature is however not equally visible at later times. 

A second important conclusion that one can draw from \autoref{CMZEvolution} is that the high-density regions formed as a consequence of gas collisions do not always evolve as gravitationally bound structures. We can see this by following the evolution of a gas cloud (indicated with a green circle in the $\alpha_\mathrm{vir}$, $L_\mathrm{HCN}$ and $T_\mathrm{B,HCN}$ maps) from $t=505$~Myr to $t=507$~Myr. During its trajectory, the cloud emission luminosity becomes weaker and weaker, going from $\sim 10^{28}$~erg~s$^{-1}$ at $t=505$~Myr, to a few times $10^{26}$~erg~s$^{-1}$ at $t=507$~Myr. This decrease in luminosity is the consequence of the cloud evaporating over time.
Indeed, although it has a high initial density, $n_\mathrm{H}\sim 7 \times10^{3}$~cm$^{-3}$ at $t=505$~Myr, the cloud is borderline gravitationally bound, $\alpha_\mathrm{vir}\sim 0.5-1$. We remind readers that $\alpha_\mathrm{vir} = 1$ for a gas cloud in virial equilibrium between turbulent motions and gravity. The cloud density is therefore not high enough to gravitationally confine the cloud against turbulent motions. Densities larger than $10^{4}$~cm$^{-3}$ are required for the gravitational collapse to start. An example is provided by the cloud highlighted with a black circle at $t  = 509$~Myr and $t  = 510$~Myr, whose virial parameter is $\sim 0.1-0.2$. The high density required to overcome the high level of turbulence explains the low level of star formation the CMZ ring is undergoing in this temporal range of the simulation. Gas at densities of $\gtrsim 10^3$~cm$^{-3}$ elsewhere in the Galaxy is almost all self-gravitating and thus star-forming, but this is not true in the much more turbulent environment of the CMZ.

\section{Discussion}
\label{Comparison}

\subsection{Comparison with the observations}
\label{Comparison with the observations}

In this Section, we use the results of our post-processed simulation when the SFR is near the minimum to interpret and spatially locate the observed features discussed in \autoref{Data}. As noted above, we cannot do a one-to-one comparison because we have not tuned the global potential to exactly reproduce the observed ring size. Moreover, we have seen that much of the dense gas structure is highly-time dependent, and so even if we matched the global size, we would not expect to reproduce every feature. Instead, our goal is to identify analogous structures in the simulated and real CMZ, and use these to interpret the observations.

Our synthetic data have shown that the dense CMZ ring produces two extended and almost parallel features in the longitude-velocity diagram. To a different extent, these two features are traced both from atomic gas and molecular gas emission. We identify these two features with Arm I and Arm II observed in the present-day CMZ. Arm I would be associated to the ring side far from the Sun, while Arm II would be associated to the ring side near the Sun. This allows us to locate Sgr C and the $20\,\kms$ and $50 \,\kms$ clouds on the near side of the ring, because, as we have seen in \autoref{Data}, their emission is kinematically associated to the Arm II emission. This is in agreement with the geometry proposed by \citet{Henshaw+16}, who argued that the $20\,\kms$ and $50 \,\kms$ clouds are situated on the near side of the CMZ because they appear to correspond to absorptions features at 70~$\mathrm{\mu}$m \citep{Molinari+11}. 

Given their alignment in longitude and velocity with Arm I, we might deduce that Sgr B2 and the dust ridge molecular clouds are a continuation of Arm I, and, therefore, located on the far side of the CMZ. In this sense, their location on the high longitude and velocity side of Arm I would be in agreement with what observed in \autoref{LowSF} and \autoref{DenseGas}, i.e. that most of the molecular gas emission associated to the far side of the ring comes from the region at higher longitudes as a consequence of cloud-cloud collision at the dust lane-ring interception. The hypothesis that Sgr B2 is the result of cloud-cloud collisions was already proposed in the past to explain its complex kinematic structure
with multiple velocity components \citep[e.g.][]{Hasegawa+94, Sato+00}. Observational studies have shown that Sgr B2 is undergoing intense star formation activity \citep[e.g.][]{McGrath+04,DePree+14}, suggesting that the collision has involved a very dense, perhaps already star-forming, cloud. On the other side, the dust ridge molecular clouds present little or no signs of star formation activity, although they have both high masses and low temperatures \citep[e.g.][]{Lis+01, Immer+12}. The dust ridge clouds might be identified with the weakly-bound molecular clouds discussed in \autoref{DenseGas} (see also the cloud highlighted with a green circle in \autoref{CMZEvolution}), which are supported against collapse by their high levels of internal turbulence.

There are two main observational constraints that are difficult to reconcile with our picture of Sgr B2 and the dust ridge molecular clouds as being located on the far side of the CMZ. First, the dust ridge clouds appear to correspond to absorption features at 70~$\mathrm{\mu}$m, much like the $20\,\kms$ and $50 \,\kms$ clouds. This would place the cloud on the near side of the CMZ, based on the assumption that clouds behind the Galactic Center are obscured from the CMZ IR emission. 
While this is suggestive, and is certainly a concern for our interpretation, we also caution that we have little information about the 3D dust and IR source distribution in the inner $1^{\circ}$ from the Galactic Center, and thus inferences about near-side versus far-side based on the presence of IR absorption features are necessarily tentative.
A second constraint is given by the observed proper motion of two water masers in Sgr B2 that indicate that this region is moving towards higher longitudes \citep{Reid+09}. However, the small number of observed sources implies large uncertainties, especially because water masers are often associated with high-velocity outflows from young stellar objects and, in this case, they might not be representative of the overall motion of Sgr B2. If a proper motion towards positive longitudes were proven true, our interpretation that Sgr B2 is rotating towards lower longitudes in the far side of the ring would certainly be disproved. 
However, in this case our interpretation of Arm I and II as the two sides of the ring could still be consistent with the observational constraints if we were to assume that Sgr B2 and the dust ridge clouds are detached from the bulk of the gas rotating on $x_2$ orbits by stellar feedback. Such detached clouds occur regularly in our simulations in the aftermath of periods of high star formation activity, when feedback locally destroys the ring-like gas distribution (see \autoref{HighSF}).

We do not identify structures with features similar to those of Arm III and Complex $1.\!\!^{\circ}3$ in our post-processed CMZ. In the \hi\ and CO longitude-velocity diagrams, we have identified the low-inclination gas stream at negative longitude as the inner low-density ring at $R\sim100$~pc. This might be compared with Arm III, which presents a similar structure in the ($l,v$) plot, even though located at positive velocities. However, there are two main inconsistencies between the feature in our simulation and Arm III. The first one is that we do not observe the inner ring in the NH$_3$ and HCN maps, since its density is not high enough ($n_\mathrm{H} \lesssim 10^3$~cm$^{-3}$) to be detected through high-density gas tracers. This inconsistency would be overcome if we post-processed the simulation with a model based on a less intense radiation field, which might be reasonable since no star formation occurs in the inner ring (see \autoref{RadFields} and \aref{append}). The main inconsistency with the observation is however that, unlike Arm III, the inner ring in not inclined with respect to the Galactic plane in the ($l,b$) diagram. Concerning Complex $1.\!\!^{\circ}3$, this has been suggested to be the accretion site of material on the CMZ. In our model, gas accretes onto the CMZ through the dust lanes. As before, the dust lane region approaching the ring is not visible through high-density gas tracers in our post-processed CMZ. However, they could be detected assuming a less intense radiation field, as we show in \aref{append}. We cannot further speculate on the possible connection of Complex $1.\!\!^{\circ}3$ with the dust lanes, since we have restricted our analysis to the ring region only, without post-processing the outer parts of the Galaxy.

Finally, from the analysis of the synthetic ($l,b$) maps, we have seen that, similarly to what is observed in the present-day CMZ, clumps of dense gas are distributed within a few parsecs above and below the plane, suggesting that their orbits are not perfectly aligned with the Galactic plane. However, the observed CMZ presents a larger extension in latitude compared to the simulation. For example, Sgr B2 extends over $-0.\!\!^{\circ}15 \leq b \leq 0.\!\!^{\circ}1$, while in the simulation the densest structures are located at $\vert b \vert \leq 0.\!\!^{\circ}05$.

\subsection{Comparison with previous works}

\begin{figure*}
\includegraphics[width=\textwidth]{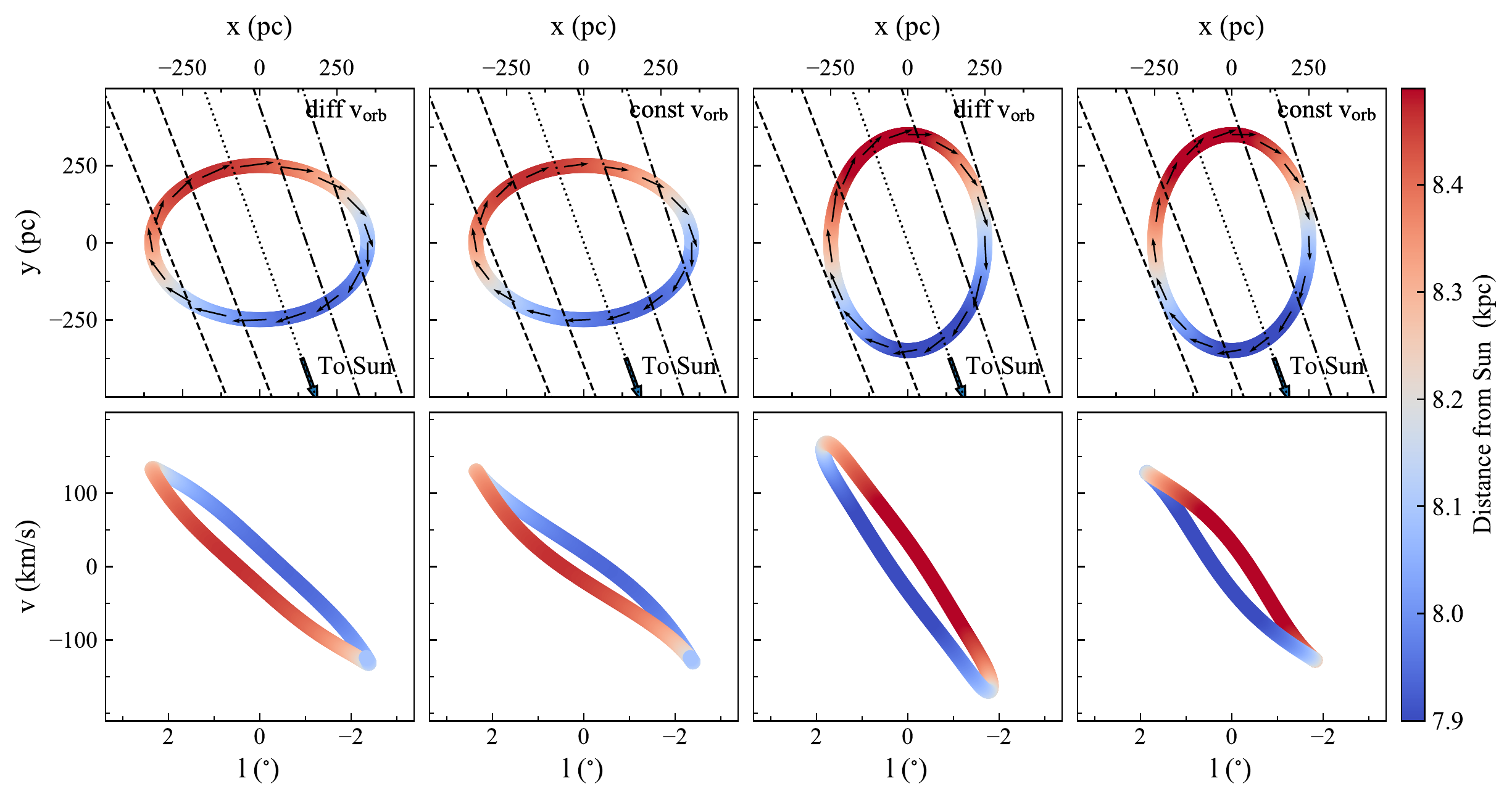}
\caption{Models of possible ring configurations in the ($x,y$) (\textit{top panels}) and ($l,v$) (\textit{bottom panels}) space. In all the models, the bar major axis is always aligned with the axis $x = 0$ and the angle between the bar major axis and the Sun-Galactic Centre line is $20^{\circ}$. In the \textit{first column} from the left, the ring minor axis is aligned with the bar major axis and the orbital velocity, $v_\mathrm{orb}$, varies across the orbit. This configuration corresponds to the one produced in our simulation. In the \textit{second column}, the ring minor axis is aligned with the bar major axis and the orbital velocity is constant. In the \textit{third column}, the ring major axis is aligned with the bar major axis and the orbital velocity is variable. In the \textit{forth column}, the ring major axis is aligned with the bar major axis and the orbital velocity is constant. The orbital velocity is shown in the top panels with vectors, whose length is proportional to the velocity magnitude. Different colors indicate the distance of each point from the Sun.}
\label{RingModels}
\end{figure*}

Different models have been proposed to interpret the 3D gas distribution within the CMZ. In agreement with our model, the ``elliptical orbit'' model proposed by \citet{Molinari+11} predicts that the CMZ represents an elliptical ring of gas moving on $x_2$ orbits. In this model, all the prominent molecular clouds (see \autoref{Data}) are located on the near side of the ring, to be consistent with the assumption that the dust ridge clouds and the $20\,\kms$ and $50\,\kms$ clouds appear as absorption features at 70~$\mu$m because they lie in the front side of the ring. The main difference from our model is that \citet{Molinari+11} assumed a constant orbital velocity, while in our simulation the orbital velocity varies depending on the orbital phase angle. In particular, it increases in proximity of the minor axis, and decreases in proximity of the major axis. 

To give an idea of how different configurations of rotational velocity affects the gas distribution in ($l,v$) space, in \autoref{RingModels} we show simplified ring models in both ($x,y$) and ($l,v$) space. In the first column from the left, we model an orbit with variable rotational velocity ($v_\mathrm{rot} \propto a/R$, where $a$ is the size of the minor semi-axis), representative of our simulation. In the second column, we model an orbit with constant orbital velocity, representative of the \citeauthor{Molinari+11} model. In both the examples, the angle between the bar major axis and the Sun-Galactic Centre line is $20^{\circ}$. The model with constant orbital velocity still produces two elongated features in the ($l,v$) map, corresponding to the near and far side of the ring. However, their profiles present more asymmetries in velocity compared to the model with variable velocity. We highlight that, given the alignment of the minor axis of the $x_2$ orbit with the bar major axis, the stream at higher line-of-sight velocities always corresponds to the near side of the ring, regardless of the velocity field. The first and second top panels of \autoref{RingModels} clearly show that the projection of the orbital velocity on the line of sight is larger for gas on the near side of the ring. The opposite would be true if the major axis of the elliptical orbit were aligned with the bar major axis (see third and forth column of \autoref{RingModels}). This configuration would be however inconsistent with the dynamical studies predicting that the barred gravitational potential produces a family of closed $x_2$ orbits perpendicular to the bar \citep[e.g.][]{Binney+91, Athanassoula92}. 

The ``elliptical orbit'' model  proposed by \citet{Molinari+11} was questioned by \citet{Kruijssen+15} and \citet{Henshaw+16}. 
\citet{Kruijssen+15} identified a significant discontinuity between the dust ridge and the $20\,\kms$ and $50\,\kms$ clouds in the ($l,v$) space, suggesting that they cannot be part of the same coherent gas stream. \citet{Henshaw+16} confirmed this finding, showing that SgrB2 and the dust ridge clouds are not physically connected to Arm II (see \autoref{SCOUSE}). This is also the case in our interpretation, where SgrB2 and the dust ridge clouds are connected to Arm I. 
Moreover, \citet{Kruijssen+15} argued that the \citeauthor{Molinari+11} model is unable to explain the presence of a third stream of gas identified in the dust ridge region (see \autoref{Data}). This led \citet{Kruijssen+15} to develop a new model, the ``open stream'' model, where all the coherent features of gas identified in the observations lie along a single open ballistic orbit, rather than along a closed orbit. The model was obtained by integrating orbits in an empirically-constrained Milky-Way potential \citep{Launhardt+02} and making the fit to the observed gas distribution and kinematics. 
In agreement with the observational constraints discussed in \autoref{Comparison with the observations} (detection of dust ridge, $20\,\kms$ and $50\,\kms$ clouds in absorption at 70~$\micron$ and  proper motion of Sgr B2), the ``open stream'' model locates all the prominent molecular clouds in front of the Galactic Center, even though on different segments of the orbit.
Although this model appears to be the most successful proposed so far in reproducing the kinematics of dense gas \citep{Henshaw+16}, it does not attempt to capture the origin of the CMZ and its connection with the gas dynamics on larger scales. In our simulation, the Galactic bar induces gas to flow towards the Galactic Center and that settles into a ring. We generally identify only two streams of gas in the position-longitude plot, corresponding to the two sides of the ring. However, it may happen that gas clouds depart from their $x_2$ orbit due to dissipative processes, producing structures not parallel to the two main streams and making the overall geometry more asymmetric. 

A third model proposed to explain the gas distribution within the CMZ is the ``spiral arms'' model \citep{Sofue95, Sawada+04}. This model assumes that the CMZ is dominated by two spiral arms, where the arm near the Sun corresponds to Arm I, while the arm farther from the Sun corresponds to Arm II. \citet{Ridley+17} re-proposed this model based on the analysis of their Milky Way-like simulation. However, unlike \citet{Sofue95} and \citet{Sawada+04}, they identified Arm I as the far spiral arm and Arm II as the near spiral arm. While \citet{Ridley+17} adopted the same gravitational potential as we use in our work, there are nonetheless several differences between the two simulations that might explain the different gas configuration in the CMZ. In \citet{Ridley+17}, self-gravity and stellar feedback are not included and, more important, gas is assumed to be isothermal with an effective sound speed of $10\,\kms$. \citet{Sormani+18} showed that the gas thermal pressure plays an important role in regulating the gas distribution within the CMZ.

We finally discuss our results in relation to the hypothesis that Sgr B2 and the dust ridge clouds represent a time-sequence of star-forming clouds \citep{Longmore+13b}. Observational studies of the dust ridge have indeed shown signs of increasing star formation activity with increasing the Galactic longitude \citep{Lis+01}. Based on the ``open streams'' predictions that the dust ridge clouds are moving towards higher longitudes, \citet{Longmore+13b} proposed that star formation might be triggered by tidal compression during the passage of gas near the minimum of the Galactic potential. This would explain why Sgr B2, located at high longitudes, exhibits high level of star formation, while the Brick, which has just experienced its passage near the Sgr A*, shows no sign of star formation. In our simulation, we do not see evidence of gas compression in proximity of the Galactic Center. Gas is mainly compressed in the regions where the dust lanes collide with the ring, i.e. close to major axis of the orbit. We have seen that, when the CMZ is in a quiescent period of star formation, many observed gas clouds are transient structures that are weakly bound or completely unbound, and may therefore slowly evaporate as they orbit. Therefore, according to our simulation, a time-sequence of star-forming clouds is more likely to form as a consequence of gas evaporation as the clouds move away from the dust lanes, rather than due to their passage near the Galactic Center. However, we do warn that the strength of the tidal perturbations that clouds experience is a function of the orbits they take, and thus of the global potential. It is conceivable that tidal shocking might be more important for clouds in a ring that is closer to Sgr A* than that in our simulation. For example, recent simulations of individual molecular clouds orbiting in the Milky-Way potential empirically-constrained by \citet{Launhardt+02} have shown that at $R\sim 100$ pc, tidal perturbations can be effective in triggering the cloud gravitational collapse \citep{Dale+19}.

Future observations will be valuable to distinguish among all the proposed models. We have seen that the \hi\ emission is uniform across the ring and able to trace its entire structure. The ongoing GASKAP survey \citep{Dickey+13} will provide high resolution and high sensitivity \hi\  observations of the Galactic Center, which might be useful to understand the large-scale gas distribution and dynamics in the CMZ. On the other side, ALMA observations of a variety of molecular transitions will give unprecedented details of individual molecular cloud properties.

\subsection{Predictions for extragalactic nuclear regions}

In \citetalias{Armillotta+19}, we suggested that the extreme environmental conditions typical of nuclear regions of barred spiral galaxies cause the star formation activity to go through burst/quench cycles, rather than reaching a steady-state equilibrium (see also \autoref{SimOverview} and \autoref{LowResSim}). In the highly-dense nuclear regions, the dynamical times of star formation are almost one order of magnitude smaller than the time required for stellar feedback to be effective. This might lead to a sudden gravitational collapse of the gas and a burst of star formation, that is suppressed only when stellar feedback significantly raises the local level of turbulence. This scenario is supported by observational studies showing that the nuclear regions of star-forming galaxies exhibit a much wider range of depletion times (ratio of the gas surface density to the star formation surface density) than the outer discs \citep[variations over $\sim 1$~dex,][]{Leroy+13, Utomo+18}.

In our model, variations in the depletion time are the result of cyclic variations in the star formation rate per unit gas mass, which is associated with a morphological transition in the gas distribution. This represents an observationally-testable prediction.
When the CMZ is near the minimum of a star formation cycle (SFR~$\sim 0.1-0.2\,\moyr$), most of the gas lies in a ring-like structure, with the regions at higher density generally located near the intersection with the dust lanes. By contrast, when the CMZ is near the maximum of a star formation cycle (SFR~$\sim 1-2\,\moyr$), the gas distribution is much less uniform, with a considerable part of the gas lying in star-forming molecular clouds, and a smaller part of it distributed in fragments of the pre-existing ring destroyed by supernova feedback. This reflects in a less coherent kinematical structure compared to the case with low star formation, which is detectable both through atomic and molecular gas emission. 

The CMZ of our Galaxy only provides a single snapshot of the star formation cycle, and thus cannot provide a direct test of this prediction. However, detailed high-resolution observations of extragalactic CMZs can probe a diverse range of nuclear environments and star-forming states. Examining \autoref{Maps_LowSFR} and \autoref{Maps_HighSFR}, the morphological difference is clearly visible in a variety of molecular tracers. Thus high-resolution maps of nuclear regions represent the best method available to test this prediction. Recently, Callanan et al.~(submitted) have obtained ALMA observations in the innermost 500~pc of the barred galaxy M83  at a resolution scale of $\sim10$~pc. The nuclear regions of M83 presents gas properties (e.g. gas content, velocity dispersion, metallicity) similar to those of the Milky Way's CMZ \citep[see also][]{Israel01, Gazak+14}. However, while our CMZ is currently producing stars at a rate lower than what expected from density-dependent star formation laws (SFR~$\sim 0.1\, \moyr$), the centre of M83 is actively star forming \citep[SFR~$\sim2.5\,\moyr$,][]{Muraoka+07}. The factor of 25 difference in SFR might be explained by the fact that the CMZs of the Milky Way and M83 are undergoing different phases of their star formation cycles. In this scenario, the CMZ of the Milky Way would be currently near a minimum of a burst/quench cycle, while the CMZ of M83 would be near a maximum. Thus the gas morphology in the centre of M83, and its level of clumping compared to that seen in the Milky Way, represents a useful test of our model. However, given the difficulties of comparing our face-on view of M83 to our edge-on view of the Milky Way, a much cleaner test would be to compare M83 to the morphology of another external galaxy with similar bulk properties, but with a present-day nuclear star formation rate more similar to that of the Milky Way than to M83.

\section{Conclusions}
\label{Conclusions}

The CMZ is an unique environment in our Galaxy characterized by extreme gas properties. Despite the gas densities orders of magnitude larger than those measured in the disc, the SFR is modest. Moreover, rather than being distributed relatively smoothly with radius as is the case elsewhere in the disc, gas in the CMZ is concentrated in a dense ring or stream structure at a limited range of galactocentric radii. In \citetalias{Armillotta+19}, we investigated the gas cycle and star formation history of the CMZ through a detailed hydrodynamical simulation of the inner 4.5~kpc of a Milky Way-like galaxy. One of the main finding of \citetalias{Armillotta+19} is that star formation activity in the CMZ goes through oscillatory burst/quench cycles, with characteristic variability-time of $\sim 50$~Myr mainly driven by stellar feedback instabilities. 

In this work, we re-run part of the simulation analysed in \citetalias{Armillotta+19} at higher resolution. The goal is to perform a detailed study of the gas distribution and kinematics in the CMZ region. We run the high-resolution simulation for 50 Myr, during which the SFR increases by more than one order of magnitude. We then carry out a detailed chemical and observational post-process of the simulated CMZ in order to produce synthetic data cubes and maps of \hi, CO~($J=1-0$), NH$_\mathrm{3}$~(1,1) and HCN~($J=1-0$) emission. The major findings of our work are as follows:
\begin{itemize}
\item When the CMZ is near a minimum of its star formation activity, most of the gas lies in a dense elliptical ring. This translates into two elongated and nearly parallel streams of gas in the longitude-velocity map. The stream at higher velocities corresponds to the near side of the ring, while the stream at lower velocities corresponds to the rear side of the ring. Gas moves from positive longitudes and velocities to negative longitudes and velocities in the near ring side, and from negative longitudes and velocities to positive longitudes and velocities in the far ring side. While the \hi\ distribution is nearly uniform across the ring, the molecular gas emission is asymmetric and predominantly clumpy. Generally, most of the emission comes from the high longitude and velocity (in absolute value) regions where gas orbiting within the ring collides with dust lane material.
\item Within the ring, shocks due to colliding flows and clouds readily produce dense regions ($n_\mathrm{H} \gg 10^3$~cm$^{-3}$) that can be observed in tracers such as HCN and NH$_3$. However, the newly formed high-density regions do not necessarily evolve as gravitationally bound structures. Densities larger than $10^4$~cm$^{-3}$ are required for the gravitational collapse to start. Gas clouds that are not sufficiently dense slowly evaporates during their trajectory within the ring. The presence of large quantities of dense but unbound gas explains the low star formation rate per unit mass in the present-day CMZ. It also suggests that at least some of the clouds that are not currently star-forming may never become so, and may instead evaporate due to hydrodynamic stripping.
\item The gas distribution and kinematics appear much less uniform and coherent when the CMZ is actively star-forming. Supernova feedback locally increases the level of turbulence, thus fragmenting the dense ring. Most of the gas lies in star-forming dense clouds rotating nearly in the same position of the pre-existing ring. The two elongated streams of gas in the longitude-velocity plot are replaced by ridges of gas with little extension in longitude and velocity in the \hi\ map and bright clouds in the molecular emission maps. In extragalactic observations of CMZs similar to that of the Milky Way, but near the maximum of their star formation cycle, this difference in morphology should be detectable in high-resolution molecular line maps.
\end{itemize}

We identify some analogies between the observed features of the Milky Way's CMZ and the gas distribution predicted by our model when the CMZ is near the minimum of the SFR. In agreement with our findings, the molecular gas distribution of the present-day CMZ is highly asymmetric and dominated by two extended parallel streams in the longitude-velocity space. However, our model would locate Sgr B2 and the dust ridge clouds beyond the Galactic Center, in contrast with some observational evidence suggesting that they should lie in front of the Galactic Center. We conclude stressing that high-resolution observations over a variety of neutral and molecular gas transitions both in the Milky Way and  extragalactic CMZs are still required to test the validity of our model against other possible interpretations. 

\section*{Acknowledgements}

The authors thank the referee, Diederik Kruijssen, for his comments that have improved the clarity of this work. They also thank Adam Ginsburg and Steven Longmore for useful suggestions and Jonathan Henshaw for providing the fit to the NH$_3$ data made with \textsc{SCOUSE}.
Simulations were performed on the Raijin supercomputer at the National Computational Infrastructure (NCI), which is supported by the Australian Government, through grant jh2. LA and MRK acknowledge support from the Australian Research Council's \textit{Discovery Projects} and \textit{Future Fellowships} funding schemes, awards DP190101258 and FT180100375. EDT acknowledges the support of the Australian Research Council (ARC) through grant DP160100723. 

%%%%%%%%%%%%%%%%%%%%%%%%%%%%%%%%%%%%%%%%%%%%%%%%%%



\bibliographystyle{mnras}
\bibliography{biblio}
    
\appendix
\section{Radiation field models}
\label{append}

\begin{figure*}
\includegraphics[width=\textwidth]{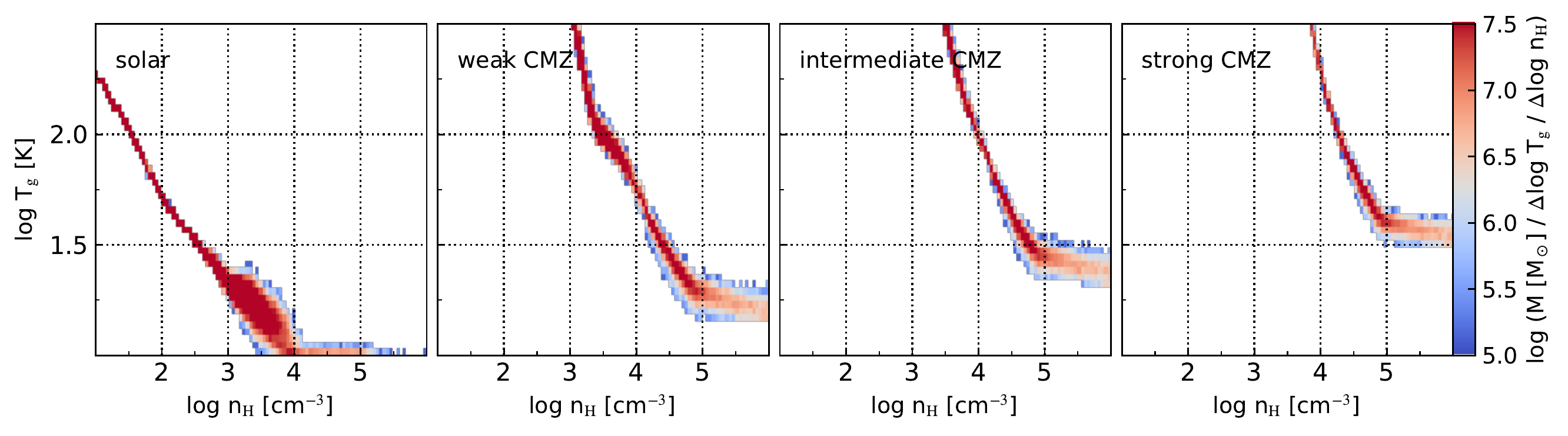}
\caption{Temperature-density phase diagram of the CMZ as calculated with \textsc{DESPOTIC} for four different radiation field models, \textit{Solar} ($\chi = 1\,G_0$ and $\zeta = 10^{-16}$~s$^{-1}$, \textit{first panel}), \textit{weak} CMZ ($\chi = 10^2 \,G_0$ and $\zeta = 10^{-15}$~s$^{-1}$, \textit{second panel}), \textit{intermediate} CMZ ($\chi = 10^{2.5}\,G_0$ and $\zeta = 10^{-14.5}$~s$^{-1}$, \textit{third panel}) and \textit{strong} CMZ ($\chi = 10^3\,G_0$ and $\zeta = 10^{-14}$~s$^{-1}$, \textit{fourth panel}). The diagram is computed as a two-dimensional histogram showing the mass of gas particles within each logarithmic bin, normalised by the bin area. The temperature-density distributions have been calculated at $t = 505$~Myr.
}
\label{Temperatures}
\end{figure*}    

\begin{figure*}
\includegraphics[width=\textwidth]{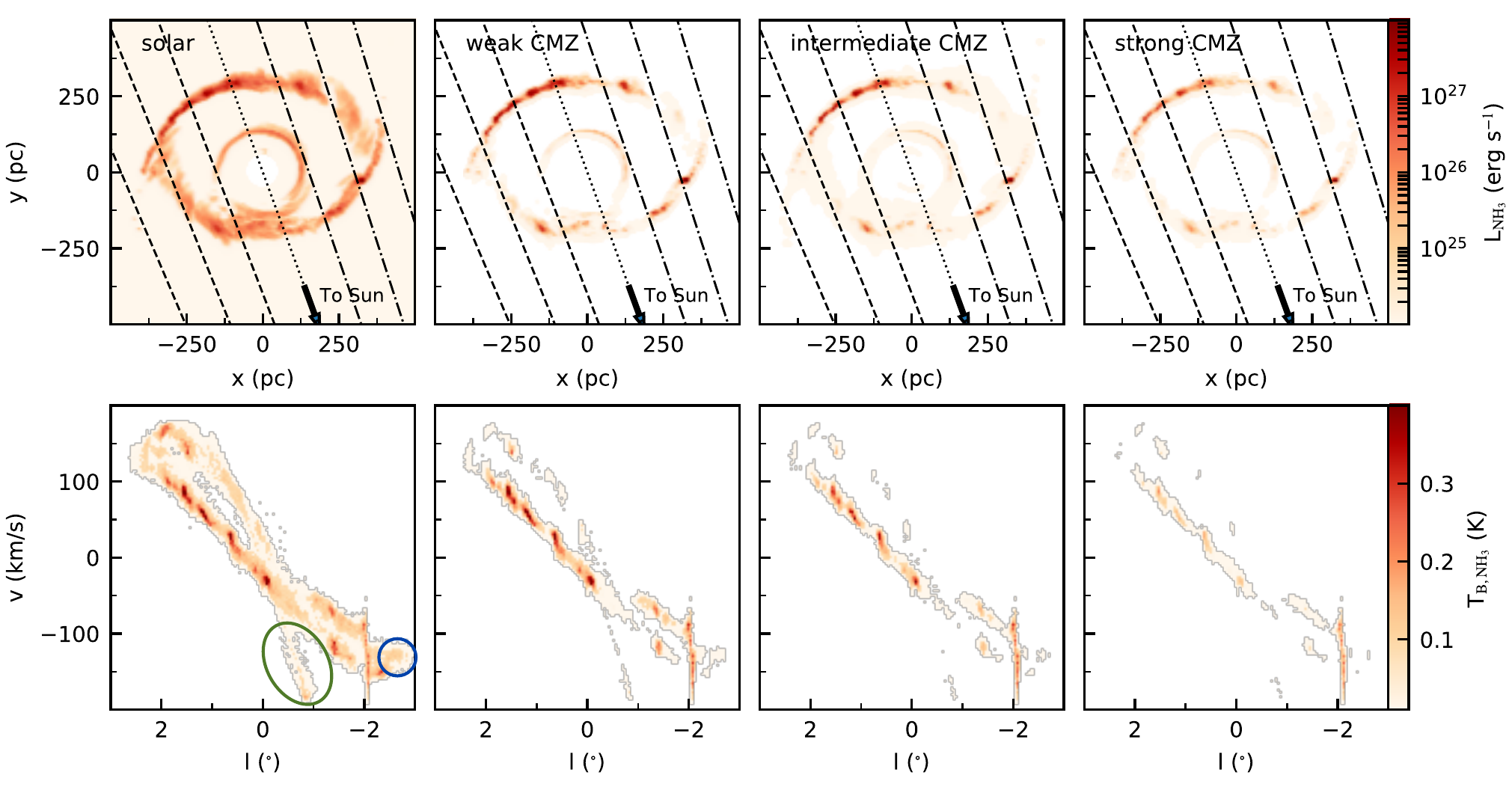}
\caption{Cartesian distribution of the NH$_3$~(1,1) luminosity (\textit{upper panels}) and NH$_3$~(1,1) brightness temperature distribution in the longitude-velocity diagram (\textit{lower panels}) obtained assuming four different radiation field models, \textit{Solar} (\textit{first column}), \textit{weak} CMZ (\textit{second column}), \textit{intermediate} CMZ (\textit{third column}) and \textit{strong} CMZ (\textit{fourth column}). Coloured contours denote relevant structures discussed in the main text. The snapshots have been taken at $t=505$~Myr.}
\label{RadFields}
\end{figure*}    
    
To establish which of the four radiation models (see \autoref{Postprocess}) is more representative of the present-day CMZ, we perform a few comparisons between the gas properties predicted by the four models and those measured in the CMZ. First of all, we derive the molecular gas mass, $M_\mathrm{mol} = 2 m_\mathrm{H} n_\mathrm{H} X_{\rm H_2}/ \mu_{H}$, where $\mu_{He} = 1.36$ is the helium correction based on cosmological abundances, and the H$_2$ abundance, $X_\mathrm{H_2}$, is an output of \textsc{DESPOTIC}. We find that the molecular mass is roughly constant with time, it varies by less than a factor 1.5 for all the models. Among them, the \textit{Solar} model predicts $M_\mathrm{mol} \sim 6\times10^7\,\mo$,  while the three CMZ models predict $M_\mathrm{mol} \sim 1.5-2.5\times10^7\,\mo$. We point out that the \textit{Solar} model and the CMZ model predictions are consistent with the upper and lower limit of the observationally-inferred masses ($M_\mathrm{CMZ} \sim 2-7\times10^7\, \mo$), respectively. However, the latter are measured for the entire inner $500$~pc region of our Galaxy. The molecular mass in the central $2^{\circ}$ of the Milky Way -- which is the region that we compare with the dense ring formed in the simulation -- is $\sim 1.8\times10^7\, \mo$ \citep{Longmore+13a}, in agreement with the predictions of the CMZ radiation models. However, we note that the molecular mass in the CMZ is often inferred using CO~($J=1-0$) emission measurements \citep[e.g.][]{Ferriere+07}. Translating from $L_\mathrm{CO}$ to $M_\mathrm{mol}$ relies on the choice of the CO-to-H$_2$ conversion factor, $\alpha_\mathrm{CO} \equiv M_\mathrm{mol}/L_\mathrm{CO}$. Since our models independently predict the total CO luminosity and the H$_2$ mass, we can evaluate $\alpha_{\rm CO}$ in our models, and compare to inferences of this same quantity in the observed CMZ obtained by comparing dust and CO maps. We find that the $\alpha_{\rm CO}$ oscillates around $1.5\,\mo$~(K~$\kms$~pc$^2$)$^{-1}$ for all the CMZ models, in agreement with the value inferred in the present-day CMZ  \citep{Kruijssen+14}.

As a second test, in \autoref{Temperatures} we evaluate the gas temperature distribution as calculated with \textsc{DESPOTIC} as a function of $n_\mathrm{H}$. The \textit{Solar} model (first panel) predicts gas temperatures well below the actual gas temperature produced in the simulation, especially at $n_\mathrm{H} \lesssim 10^2$~cm$^{-3}$. This is because the photoelectric heating rate (\autoref{Code}) used in the simulation is chosen to mimic an ISRF significantly stronger than that found in the Solar neighbourhood, and thus the transition from warm neutral medium (WNM, $T\sim10^4$~K) to cold neutral medium (CNM, $T\lesssim100$~K) occurs at higher gas densities in the simulation than in the post-processed table for the \textit{Solar} case. The three CMZ radiation field models all produce similar WNM / CNM transitions, but differ in the temperatures they predict in dense and shielded gas at $n_\mathrm{H} \sim 10^4-10^5$~cm$^{-3}$. In particular, the temperatures predicted by the $weak$ model, $T_\mathrm{g} \sim 15-60$~K (second panel), are in agreement with those characteristic of the cold molecular component \citep[$T \sim 25-50$~K,][]{Krieger+17}, while the temperatures predicted by the $strong$ model, $T_\mathrm{g} \sim 40-180$~K (fourth panel), are in agreement with the observed temperatures of the warm molecular component \citep[$T \sim 60-150$~K,][]{Krieger+17}. As expected, the $intermediate$ model (third panel) produces intermediate temperatures between the two previous models, $T_\mathrm{g} \sim 25-100$~K. This result clearly suggests that the radiation field in the present-day CMZ is not constant, but instead varies based in the local star formation conditions.  Given the best match with the observations in terms of temperature, we adopt the \textit{intermediate} CMZ radiation model as the fiducial model throughout this work. 

To understand the sensitivity of our results to this choice, in \autoref{RadFields} we compare the distribution of NH$_3$~(1,1) emission predicted by the four models at $t=505$~Myr. The NH$_3$~(1,1) luminosity becomes weaker and weaker with increasing ISRF and cosmic-ray fluxes as a consequence of the different gas temperature and molecular gas mass predicted by the different models. The \textit{Solar} radiation model predicts high luminosities ($L_{\rm NH_3} >10^{26}$~erg~s$^{-1}$) across the entire ring. This allows us to clearly detect all the prominent features in the longitude-velocity map that are discussed in the main text: the two elongated parallel streams corresponding to the two sides of the ring, the ridge of molecular clouds in the far side of the ring, and also the low-density inner ring at $R\sim 100$~pc (green contour), that we have seen to be mainly composed of atomic gas under the assumption of \textit{intermediate} CMZ radiation field (see \autoref{LowSF}). On the other hand, under the assumption of \textit{strong} CMZ radiation field, the NH$_3$~(1,1) brightness temperatures are very low ($T_\mathrm{B, NH3} \lesssim 0.2$~K) even in regions with relatively-high density ($n_\mathrm{H} \lesssim10^4$~cm$^{-3}$). The blue contour in the \textit{Solar} ($l,v$) map highlights the dust lane side immediately connected to the ring. The brightness temperature associated to this region becomes a factor of 2 lower for the \textit{weak} CMZ radiation field compared to the \textit{Solar} case, and nearly zero assuming \textit{intermediate} or \textit{strong} radiation. As noted above, it is very likely that the radiation field is not constant throughout the CMZ region, but instead varies depending on local star formation activity. Therefore, we cannot exclude the possibility that regions such as the dust lanes, where little star formation occurs, might experience a less intense radiation field and produce higher brightness temperatures than those predicted in our fiducial, \textit{intermediate} case.

%%%%%%%%%%%%%%%%%%%%%%%%%%%%%%%%%%%%%%%%%%%%%%%%%%

% Don't change these lines
\bsp	% typesetting comment
\label{lastpage}
\end{document}